\documentclass[10]{article}
\usepackage[colorlinks=true, allcolors=blue]{hyperref}
 \newtheorem{theorem}{Theorem}
\usepackage[top=0.9in, bottom=0.9in, left=0.7in, right=0.7in]{geometry}
\usepackage{natbib}
\usepackage{amsmath}
\usepackage{amsfonts}
\usepackage{amssymb}
\usepackage{graphicx}%
\providecommand{\U}[1]{\protect\rule{.1in}{.1in}}

\newtheorem{proposition}[theorem]{Proposition}

\renewcommand{\textsl}[1]{{#1}}

\title{\bf Two Types of Temporal Symmetry in the Laws of Nature}

\author{A.Y. Klimenko \\ Centre For Multiscale Energy Systems \\ School of Mechanical and Mining Engineering \\ The University of Queensland \\ St. Lucia 4072, Australia \\ Email: a.klimenko@uq.edu.au   \\  \\ \textit{ Published: \textbf{ Entropy 2025, 27}(5), 466}  \\  https://doi.org/10.3390/e27050466  }

\date{}

\begin{document}
\maketitle

\abstract{This work explores the implications of assuming time symmetry and applying bridge-type, time-symmetric temporal boundary conditions to deterministic laws of nature with random components. The analysis, drawing on the works of Kolmogorov and Anderson, leads to two forms of governing equations, referred to here as symmetric and antisymmetric. These equations account for the emergence of characteristics associated with conventional thermodynamics, the arrow of time, and a form of antecedent causality. The directional properties of time arise from the mathematical structure of Markov bridges, without requiring any postulates that impose a preferred direction of time. \\ \\ {\bf Keywords:} Temporal symmetry, Markov diffusion bridge, Arrow of time, Entropy increase }



\section{Introduction}

The \textit{arrow of time} remains one of the most evident and most guarded
secrets of nature. In this context, major theories and models can be divided
into two groups: (1) \textit{time-directional,} which recognises inequivalence
of the directions of time, and (2) \textit{time-symmetric, }which implies that
the directions of time are conceptually equivalent. Thermodynamics, chemical
kinetics, and viscous and diffusive fluid mechanics belong to the first group,
which explicitly incorporate time-directional, irreversible evolutions. For
example, the second law of thermodynamics states that the entropy of an
isolated system must not decrease as time moves forward. In contrast,
classical mechanics, relativistic mechanics, unitary quantum mechanics, and
electromagnetism are fundamentally time-reversible. In these theories, the
directionality of time is conventionally introduced through \textit{antecedent
causality}. While causality reflects a general scientific principle of
uncovering deep underlying connections between events or processes, antecedent
causality specifically emphasises the temporal precedence of causes over
effects. Although there are various interpretations of causality
\cite{Hume1748,Russell2009,Reichenbach1971,SE-Caus}, our understanding of
antecedent causality is focused on the conceptual and practical preference for
specifying \textit{initial} rather than \textit{final} conditions. This
preference discriminates the directions of time (i.e. introduces the
\textit{arrow of time}) directly or with common corollaries such as:
dependencies appear after but not before interactions, or that an intervention
is followed by---but not preceded by---its effect. The combination of
time-symmetric theories with antecedent causality often allows the
demonstration of entropy increase, as is done in Boltzmann's H-theorem
\cite{Boltzmann-book} or Zurek's decoherence theory \cite{Zurek2003}.


While antecedent causality is deeply embedded in our intuition and remains
implicitly present in most theories and models, beginning with Hume
\cite{Hume1748}, Boltzmann \cite{Boltzmann-book} and Russell
\cite{Russell2009}, it is commonly accepted that antecedent causality is a
valuable intuitive tool but not a necessity of thought or the most fundamental
underlying property of nature. Starting with Boltzmann \cite{Boltzmann-book}
and Reichenbach \cite{Reichenbach1971}, many philosophers tend to define or
relate antecedent causality to the time-directional properties of the second
law. This leads to the fundamental \textit{logical circuit}: antecedent
causality is defined in relation to the entropy increase and entropy increase
is demonstrated using antecedent causality \cite{mixing2019b}.

If, for analysing the direction of time, we wish to avoid a priori
discrimination of the directions of time (which is rather difficult due to our
well-developed temporal intuition \cite{PriceBook}), the practice of
explicitly or implicitly imposing initial (and not final) conditions must give
way to \textit{time-symmetric temporal boundary conditions} jointly set in the
past and in the future. The two-state vector formalism in quantum mechanics
\cite{2S-QM2008} is a good example of a model that offers such time-symmetric
interpretation of laws of nature by representing the current state as a
combination of two quantum waves: $\left\vert \varphi\right\rangle $
propagating forward in time and satisfying some initial conditions $\left\vert
\varphi\right\rangle _{t=t_{1}}=\varphi_{1}$, and $\left\langle \psi
\right\vert $ propagating backward in time and satisfying some final
conditions $\left\langle \psi\right\vert _{t=t_{2}}=\psi_{2}$, where
$t_{2}>t_{1}.$\ The product $\langle\psi|\varphi\rangle$ is then proportional
to a probability, which somewhat resembles the structure of the probability
equations presented in this work. It should be noted that, unlike the present
approach, the two-state vector formalism --- being based on unitary
transformations --- cannot account for irreversible effects or generate an
arrow of time. \textsl{Alternatively, Drummond \cite{Drum2021} shows that the
unitary evolution of a quantum field can be reformulated as a time-symmetric
Fokker--Planck equation, allowing for both forward and backward stochastic
propagation. While Drummond's approach bears some mathematical resemblance to
our present treatment, the physical role of randomness that concerns
us---priming thermodynamic irreversibilities---is entirely different.}

The relationship between the thermodynamic arrow of time and cosmic dynamics
was promoted by Gold \cite{Gold1962}, who suggested that the expansion of the
universe is associated with increasing entropy and, therefore, that
contraction should correspond to decreasing entropy. Hawking followed this
line of reasoning, proposing similar gravitational treatments of the Big Bang
and the Big Crunch. However, his approach was critisised by Penrose, who
argued that even if the universe were to undergo a Big Crunch, it would have
high entropy, making it fundamentally different from the Big Bang
\cite{Hawking-Penrose}. Schulman \cite{Schulman1997} explained that it is not
the expansion or contraction of the universe that determines the arrow of
time, but rather the temporal boundary conditions. While it is common to
assume that the initial state of the universe --- the Big Bang --- had low
entropy, Schulman proposed that the final state --- the Big Crunch --- may
also have low entropy. \ Although Gold's and Schulman's approaches may exhibit
similar global entropy dynamics, they rest on different physical foundations
and carry distinct implications. Tamm \cite{Tamm2021} considered cosmological
consequences of having both a low-entropy Big Bang and a low-entropy Big
Crunch, although our focus here is on the thermodynamic rather than the
cosmological aspects.

The aim of this work is to examine the implications of Schulman's formulation
\cite{Schulman1973, Schulman1997}, which imposes two temporal boundary
conditions. The present model shares a number of features and some
implications with Schulman's approach. However, unlike Schulman's original
framework, we assume the intrinsic presence of \textit{randomness} in nature.
This implies that even the smallest thermodynamic systems are ubiquitously
affected by randomness, which is considered to be ultimately responsible for
the arrow of time and that entropy is an objective, observer-independent
quantity. This assumption resolves the technical difficulties associated with
simultaneously imposing two boundary conditions in deterministic systems
(which requires dynamical mixing on galactic scales), and naturally leads to
\textit{stochastic bridge} formulations. The term \textit{bridge}
conventionally refers to stochastic processes constrained at both the initial
and final times, while \textit{stochastic} denotes genuine randomness rather
than deterministic chaos. Our goal is to construct a time-symmetric stochastic
model subject to time-symmetric boundary conditions, which is nonetheless
capable of explaining irreversible behaviour in nature and, effectively,
producing the arrow of time \textsl{reflecting the influence of the nearest
temporal boundary conditions}. This model avoids postulating antecedent
causality and instead points to physical mechanisms that can give rise to
irreversible effects and exhibit features conventionally associated with
causal behaviour.

Section 2 considers the laws of Hamiltonian mechanics (classical and quantum)
and introduces two types of temporal symmetry: odd and even. Section 3
addresses random models under bridge conditions. Sections 4 and 5 explore the
physical implications of these models. Conclusions are presented in Section 6.
The Appendices cover the mathematical properties of Markov diffusion bridges
and examines temporal reversibility of stochastic processes.

\section{The laws of mechanics and determinism\label{sec2}}

An increase in entropy corresponds to a loss of information and an increase in
uncertainty, which implies randomness. If one believes that the thermodynamic
arrow of time is real, so too must randomness be real, which from a
philosophical perspective is assumed to be \textit{ontological }rather than
\textit{epistemic}. Randomness can be interpreted in two main ways:
ontological (real, true) randomness, where random events occur inherently and
are not entirely determined by prior deterministic causes, and epistemic or
subjective randomness, where the world remains deterministic while
unpredictability arises from a lack of knowledge about the system. This work
adopts the ontological interpretation of randomness, asserting that it is not
merely a reflection of incomplete knowledge, but a fundamental aspect of
nature --- inherently present everywhere and responsible for time-directional
effects. In this view, the arrow of time and entropy increase are not just
apparent properties but are intrinsically linked to the irreducible stochastic
character of the laws of nature. The ontological perspective views entropy as
a real physical quantity, rather than something that depends on subjective
interpretations (such as coarse-graining). Yet, we also recognise the
importance of deterministic laws in establishing fundamental constraints and
governing principles that shape the evolution of physical systems, restrain
randomness and impose symmetries.

Consider a system of a very high dimension
\begin{equation}
\frac{dx^{i}}{dt}=a^{i}\left(  \mathbf{x},t\right)  +b^{ik}\left(
\mathbf{x},t\right)  \xi^{k}\left(  t\right)  \label{SDE00}%
\end{equation}
reflecting the known laws of nature, where $x^{i}$ can represent generalised
coordinates, momenta or any other dynamic parameters as needed. We assume that
these laws are discrete (e.g. represented by dynamics of discrete particles)
or allow for discretisation. The Einstein summation convention applies here
and in the rest of the paper. The first term, $a^{i}\left(  \mathbf{x}%
,t\right)  $ reflects deterministic mechanical laws (including classical,
relativistic, and quantum mechanics), which are considered time-reversible
and, on their own, do not possess a time arrow. The second term involves a
genuinely random quantity (specifically, white noise), which is responsible
for time-directional effects. This term is generally assumed to be very small,
such that deterministic laws typically exhibit high precision; yet, this small
random component serves as a \textit{time primer} --- the process ultimately
responsible for the emergence of irreversibility. Dynamical mixing in complex
mechanical or quantum systems can amplify the thermodynamic consequences of
these random contributions. Nevertheless, the realistic nature of randomness
underpins the presence of such subtle effects in every minute system we may
wish to consider.

Let us put $b^{ik}=0$ and examine common properties of the deterministic laws.

\subsection{Mechanical laws}

The essence of mechanical laws is conventionally expressed by canonical
\textit{Hamiltonian equations}. Assume $\mathbf{x}=\{\mathbf{q},\mathbf{p}\},$
\ then $\mathbf{a=}\{\mathbf{\dot{q}},\mathbf{\dot{p}}\}$ and canonical
equations become \cite{LL}%
\begin{equation}
\dot{q}^{i}\overset{\text{def}}{=}\frac{dq^{i}}{dt}\mathbf{=}\frac{\partial
H}{\partial p_{i}}=a_{(q)}^{i},\ \ \ \dot{p}_{i}\overset{\text{def}}{=}%
\frac{dp_{i}}{dt}\mathbf{=-}\frac{\partial H}{\partial q^{i}}=a_{i}^{(p)}
\label{Ham1}%
\end{equation}
implying
\begin{equation}
\frac{\partial\dot{q}^{i}}{\partial q^{i}}+\frac{\partial\dot{p}_{i}}{\partial
p_{i}}=0
\end{equation}
leading to $\partial a^{i}/\partial x^{i}=0$. This condition is important to
ensure that the deterministic equations do not produce entropy and/or lose
information and, therefore, can always be reversed and solved backward in
time. Hamiltonian equations possess some time-symmetric properties:
substituting $\bar{t}=-t,$ $\bar{q}^{i}=q^{i}$ and $\bar{p}^{i}=-p^{i}$ into
(\ref{Ham1}) yields the system
\begin{equation}
\frac{d\bar{q}^{i}}{d\bar{t}}\mathbf{=}\frac{\partial H}{\partial\bar{p}_{i}%
},\ \ \ \frac{d\bar{p}^{i}}{d\bar{t}}\mathbf{=-}\frac{\partial H}{\partial
\bar{q}^{i}} \label{Ham1b}%
\end{equation}
which remains invariant, provided $H$ depends on $\mathbf{p}^{2}$ and not on
$\mathbf{p}$. While this is true in many cases, some systems (e.g. involving
rotating frames or magnetic field) have terms that are linear with respect to
$p^{i}$. In this case, invariant time reversal requires reversal of rotation
and inverse of the direction of magnetic field (or charge conjugation).

In \textit{relativistic mechanics} $t$ is replaced by proper time $\tau$ while
the structure of the Hamiltonian equations remains the same. The definition of
the proper time in relativity $d\tau^{2}=g_{\nu\mu}dx^{\nu}dx^{\mu}$ in terms
of four-vectors $dx^{\nu}$ and the metric tensor $g_{\nu\mu},$ has an
arbitrary sign $(-d\tau)^{2}=(+d\tau)^{2}$ and, therefore, relativistic
equations tend to be fully invariant with respect to the proper time reversal
$\tau->-\tau$ \cite{LL}.

\subsection{Laws of quantum mechanics}

In quantum mechanics the \textit{Hamiltonian form} of the governing equation
$i\hbar(\partial\Psi/\partial t)=\mathbb{H}\Psi$ is complex and needs to be
separated into real and imaginary parts to match our models \cite{LL}%
\begin{equation}
\left\{
\begin{array}
[c]{c}%
\dot{q}^{i}\overset{\text{def}}{=}\frac{\partial q^{i}}{\partial t}=\frac
{1}{\hbar}\left(  H_{s}^{ij}p^{j}+H_{a}^{ij}q^{j}\right)  =a_{q}^{i}\\
\dot{p}^{i}\overset{\text{def}}{=}\frac{\partial p^{i}}{\partial t}=\frac
{-1}{\hbar}\left(  H_{s}^{ij}q^{j}-H_{a}^{ij}p^{j}\right)  =a_{p}^{i}%
\end{array}
\right.  \label{Ham2}%
\end{equation}
where $\Psi=\Sigma_{j}\left(  q^{j}+ip^{j}\right)  \left\vert j\right\rangle $
and $\mathbb{H}_{s}$ is symmetric and $\mathbb{H}_{a}$ is antisymmetric
components of the Hamiltonian operator so that $\mathbb{H}=\mathbb{H}%
_{s}+i\mathbb{H}_{a}$ is Hermitian and evolution specified by (\ref{Ham2}) is
unitary. Evaluation of divergence yields
\begin{equation}
\frac{\partial\dot{q}^{i}}{\partial q^{i}}+\frac{\partial\dot{p}^{i}}{\partial
p^{i}}=H_{a}^{ii}+H_{a}^{ii}=2\operatorname{trace}\left(  \mathbb{H}%
_{a}\right)  =0
\end{equation}
again leading to $\partial a^{i}/\partial x^{i}=0$. If that the basis is
selected to that the antiunitary time reversal operator $\mathbb{T}%
=\mathbb{KU}$ is represented in this basis by complex conjugation $\mathbb{K}%
$, then temporal symmetry requires that the Hamiltonian is real $\mathbb{H}%
=\mathbb{H}_{s}$ and symmetric, while $\mathbb{H}_{a}=0$. The time reversal of
(\ref{Ham2}) yields
\begin{equation}
\left\{
\begin{array}
[c]{c}%
\frac{\partial\bar{q}^{i}}{\partial\bar{t}}=\frac{1}{\hbar}\left(  H_{s}%
^{ij}\bar{p}^{j}\right) \\
\frac{\partial\bar{p}^{i}}{\partial\bar{t}}=\frac{-1}{\hbar}\left(  H_{s}%
^{ij}\bar{q}^{j}\right)
\end{array}
\right.
\end{equation}
where $\bar{t}=-t,$ $\bar{q}^{i}=q^{i}$ and $\bar{p}^{i}=-p^{i}$.

While non-relativistic quantum mechanics considers time reversal as an
external transformation, relativistic quantum mechanics tends to possess
direct-time and reverse-time solutions, with the former associated with
particles and the latter with antiparticles. The modern view in quantum field
theory is that the fundamental laws of the universe are not completely
time-symmetric, although they are close to being time-symmetric. Specifically,
the governing equations remain exactly the same under CPT transformation, i.e.
when time reversal is combined with parity transformation (spatial inversion)
and charge conjugation (swapping matter with antimatter).

\subsection{Time reversal in deterministic equations \label{Sec23}}

Overall the basic deterministic physical laws governing the universe tend to
preserve phase space volume (are non-divergent)
\begin{equation}
\frac{\partial a^{i}}{\partial x^{i}}=0 \label{div}%
\end{equation}
making evolutions governed by these laws physically time reversible%
\begin{equation}
\text{A1) }\frac{dx^{i}}{dt}=a^{i}\left(  \mathbf{x},t\right)  \iff
\text{A2)}\frac{dx^{i}}{d\bar{t}}=\bar{a}^{i}\left(  \mathbf{x},t\right)
=-a^{i}\left(  \mathbf{x},t\right)  ,\text{ \ \ }\bar{t}=-t\text{ }
\label{A12}%
\end{equation}
While this equation is a mathematical identity, equivalence of A1 and A2 also
has a subtle physical implication: information should not be not lost and
entropy should be preserved by these equations so that they can be
equivalently solved forward in time or backward in time. Time reversal A is
reversal of the order of the events without any other changes.

We also consider another type of time reversal, requiring that the governing
laws remain the same when time is reversed, implying
\begin{equation}
\text{B1) }\frac{dx^{i}}{dt}=a^{i}\left(  \mathbf{x}\right)  \iff
\text{B2)}\frac{d\bar{x}^{i}}{d\bar{t}}=\bar{a}^{i}\left(  \mathbf{\bar{x}%
}\right)  ,\ \ \bar{a}^{i}\left(  \mathbf{x}\right)  =a^{i}\left(
\mathbf{x}\right)  ,\text{ \ \ }\bar{t}=-t \label{B12}%
\end{equation}
In this case, the parameters of the model should not explicitly depend on time
since this would immediately create asymmetry of the directions of time. Here,
$\bar{x}^{i}\neq x^{i}$ generally, but there should be a simple transformation
$\mathbf{\bar{x}=}\mathbb{F}\mathbf{x}$ (or $\mathbf{x=}\mathbb{\bar{F}%
}\mathbf{\bar{x},}$ $\mathbb{\bar{F}F}=\mathbb{I}$\ \ ), which involves
flipping signs and, possibly, some permutations. Replacing variables in B1
results in $d\mathbf{\bar{x}}/d\bar{t}=$ $-\mathbb{F}\mathbf{a}(\mathbf{x})$
since $\mathbf{a}$ transforms contravariantly. Substituting $\mathbf{x=}%
\mathbb{\bar{F}}\mathbf{\bar{x}}$ yields $-\mathbb{F}\mathbf{a}(\mathbb{\bar
{F}}\mathbf{\bar{x}})=\mathbf{\bar{a}}\left(  \mathbf{\bar{x}}\right)
=\mathbf{a}\left(  \mathbf{\bar{x}}\right)  $ to be consistent with B2.
Finally, we obtain conditions $\mathbf{a}(\mathbb{\bar{F}}\mathbf{\bar{x}%
})=-\mathbb{\bar{F}}\mathbf{a}\left(  \mathbf{\bar{x}}\right)  $ and
$\mathbf{a}\left(  \mathbb{F}\mathbf{x}\right)  =-\mathbb{F}\mathbf{a}%
(\mathbf{x})$ for consistency between B1 and B2. It is this relation that
constitutes the temporal symmetry: without $\mathbb{F}$-consistency, B1 and B2
would simply represent different models. Examples are given in our previous
analysis (\ref{Ham1}): $\mathbf{x}=\{\mathbf{q},\mathbf{p}\}$ and
$\mathbf{\bar{x}}=\{\mathbf{\bar{q}},\mathbf{\bar{p}}\}$, where $\mathbf{\bar
{q}=q}$ and $\mathbf{\bar{p}=-p}$ defines $\mathbb{F}$. The mechanical laws
generally imply existence of B-type time symmetry but require adjustments by
$\mathbb{F}$ and may need further qualifications. While B1 and B2 may
represent similar but physically different systems, A1 and A2 can usually be
seen as evolutions of the same system examined from different perspectives.
The B-type symmetry can often be approximate and swapping some variables might
be needed to make it exact or more accurate. For example, B-type CPT-invariant
reversal of time in the universe would require all its matter replaced by
antimatter (i.e. swapping the corresponding matter and antimatter states) ---
this is only conceptual but not physical possibility.

While in dynamic equations A-type transformation is reversal of time and
B-type is more related to temporal symmetries, the terms \textit{reversal,}
\textit{reversibility} and \textit{temporal symmetry} are commonly used
interchangeably: in stochastic systems the A-type reversal implies presence of
some symmetry, while B-type symmetry involves (or can involve) time reversal
and is referred to as reversibility in a number of principal works. As we need
to clearly distinguish these cases, the B-type time symmetry is referred to as
\textit{even} and the associated transformations are called \textit{symmetric}%
, while A-type symmetry is referred to as \textit{odd} and the associated
transformations as \textit{antisymmetric}. The term \textit{temporal symmetry}
and the adjective \textit{time-symmetric} (used in the Introduction) imply
general presence of symmetry without referring to a specific type. As can be
seen in the rest of the paper, both types involve reversal of time and possess
some distinct temporal symmetries --- yet, they produce different models.
Since these models are connected to both symmetry and transformations, even
and symmetric on one side, and odd and antisymmetric on the other, can be used interchangeably.

\begin{figure}[h]
\caption{Illustration of symmetric and antisymmetric time reversal.
Forward-time porcess (solid), Symmetric reverse (dashed), Antisymmetric
reverse (dotted)}%
\label{fig_1}
\begin{center}
\includegraphics[width=14cm,page=1,trim=3cm 3cm 2.5cm 6cm, clip ]{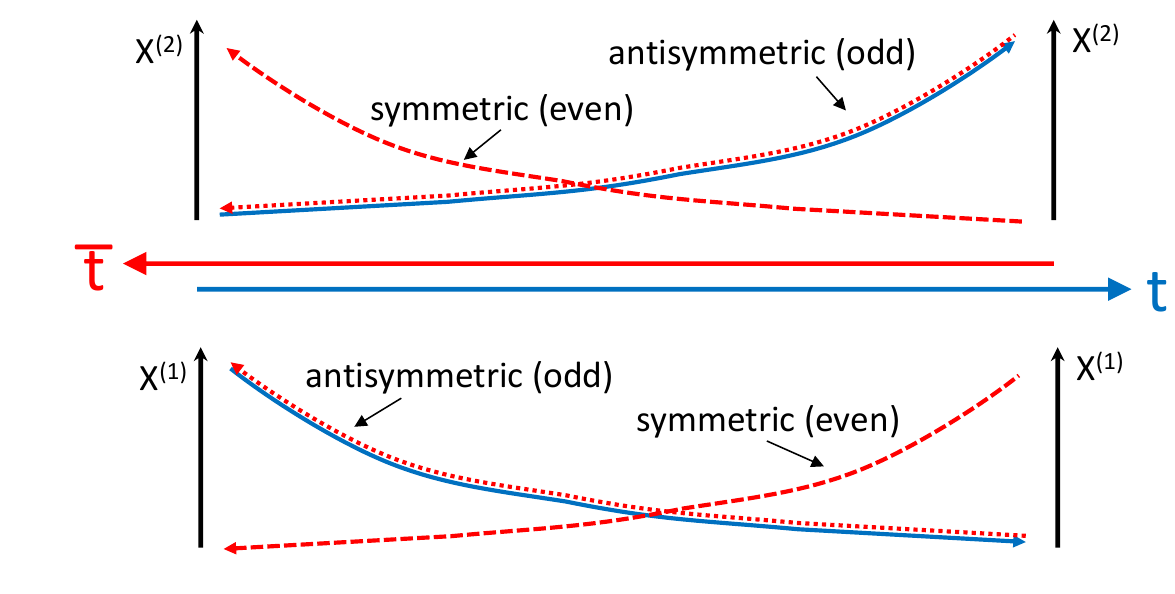}
\end{center}
\end{figure}

The following two-dimensional example, which is schematically presented in
Figure \ref{fig_1}, illustrates the difference between symmetric and
antisymmetric reversal of time. Assume $b^{i}=0$ and $a^{(1)}=-x^{(1)},$
$a^{(2)}=+x^{(2)}$ corresponding to a converging-diverging velocity field
complying with (\ref{div}) and specifying trajectories $x^{(1)}\sim e^{-t}$
and $x^{(2)}\sim e^{+t}$\ The antisymmetric time reversal $\bar{a}%
^{(1)}=+x^{(1)},$ $\bar{a}^{(2)}=-x^{(2)}$ reproduces the same trajectories
$x^{(1)}\sim e^{+\bar{t}}=e^{-t}$ and $x^{(2)}\sim e^{-\bar{t}}=e^{+t},$ so
that converging becomes diverging backward in time and vice versa. The
symmetric time reverse $\bar{a}^{(1)}=-x^{(1)},$ $\bar{a}^{(2)}=+x^{(2)}$
produces evenly similar trajectories in reversed time $x^{(1)}\sim e^{-\bar
{t}}$ and $x^{(2)}\sim e^{+\bar{t}},$ which, however, do not coincide with the
original direct-time trajectories. Note that symmetric time reversal in
conjunction with swapping $x^{(1)}\ $and $x^{(2)}$ matches the original trajectories.

\section{The random bridge models}

Our analysis requires the presence of \textit{true randomness} as, without
antecedent causality or additional assumptions, deterministic equations are
time-symmetric and fail to establish an arrow of time. Without effective
randomisation, deterministic equations tend to force a choice between initial
and final conditions, implicitly introducing antecedent causality and bringing
us back to the logical circle discussed in the introduction. Incorporating
true randomness into the model is crucial: a model without random fluctuations
(or at least some form of emergent stochasticity, such as chaotisation
associated with dynamical mixing) is rigid and could not satisfy both the
initial and final conditions at the same time. \textit{Determinism} would
force us to choose between the initial conditions set, say, at $t=t_{1}$ and
final conditions, say, set at $t=t_{2}>t_{1}$ making the model
time-directional. Having preference for initial conditions introduces
discrimination of the directions of time and is a de factor implementation of
thinking associated with antecedent causality. Randomness provides a natural
justification for imposing time-symmetric temporal boundary conditions
\begin{equation}
\mathbf{x}(t_{1})=\mathbf{x}_{1},\ \ \ \ \mathbf{x}(t_{2})=\mathbf{x}_{2}%
\end{equation}
which are commonly referred to as \textit{stochastic (random) bridges}. The
associated probability distribution $f=f(\mathbf{x},t)$ then satisfies the
corresponding initial and final conditions
\begin{equation}
\left(  f\right)  _{t=t_{1}}=\delta(\mathbf{x}-\mathbf{x}_{1}),\ \ \ \left(
f\right)  _{t=t_{2}}=\delta(\mathbf{x}-\mathbf{x}_{2})
\end{equation}

This work considers two models for the laws of nature, both of these models
involve true randomness generated by \textit{Wiener processes} and, despite
their diffusion character, possess some temporal symmetry. Due to the presence
of randomness, these models are inevitably related to some forms of the
\textit{Fokker-Planck equation}. In this context, two types of time reversal
(or reversibility) have been considered: \textit{symmetric} following
Kolmogorov \cite{Kolmogoroff1937} and \textit{antisymmetric} following
Anderson \cite{Anderson1982}.

Kolmogorov's symmetric time reversal leaves the governing equations formally
unchanged when expressed in the new time variable $\bar{t}$, even though this
re-labeling does not ensure that the system's evolution is literally reversed,
but as discussed previously the direct and reverse evolutions are evenly
similar. In contrast, Anderson's antisymmetric time reversal focuses on the
actual behavior of the system---when time is reversed, the system retraces its
exact steps, visiting each state in the reverse order. Symmetric time reversal
emphasises the formal invariance of the laws under a change of the time
coordinate, whereas antisymmetric time reversal highlights that the dynamics
themselves are truly reversible, with forward and backward trajectories
matching. As discussed in the Appendix, matching random trajectories forward
and backward in time requires to use the symmetrised \textit{Stratonovich}
interpretation \cite{Stratonovich1968} of stochastic equations instead of a
more common interpretation due to \textit{Ito} \cite{Ito1944}. Because
Kolmogorov's notion of reversibility is fundamentally a conceptual statement
about the symmetry of the transition probabilities (i.e. detailed balance), it
is not tied to any specific interpretation of stochastic differential
equations and requires only consistent definitions forward and backward in
time. In other words, Kolmogorov's reversibility does not directly reverse
stochastic trajectories and does not specifically require the symmetrised
formulation provided by the Stratonovich integral. It simply takes a
direct-time model and applies the same model backward in time; therefore, this
model can be equivalently expressed with any formulation of the stochastic
differential equations, and Ito's formulation tends to be more convenient. The
two approaches to reversing time result in the corresponding models derived in
the Appendix, schematically illustrated in Figure \ref{fig_2} and presented below.

\begin{figure}[h]
\caption{Illustration of symmetric and antisymmetric time reversals in
stochastic systems. The shaded areas reflect probability densities; the dotted
and solid arrows indicate the magnitudes of the diffusion and drift,
respectively, for the forward (blue) and reversed (red) processes. The long
arrows illustrate transition probabilities between states 1 and 2, which are
consistent with detailed balance required by Kolmogorov's reversibility
conditions.}%
\label{fig_2}
\begin{center}
\includegraphics[width=14cm,page=2,trim=1.5cm 1cm 1cm 1.7cm, clip ]{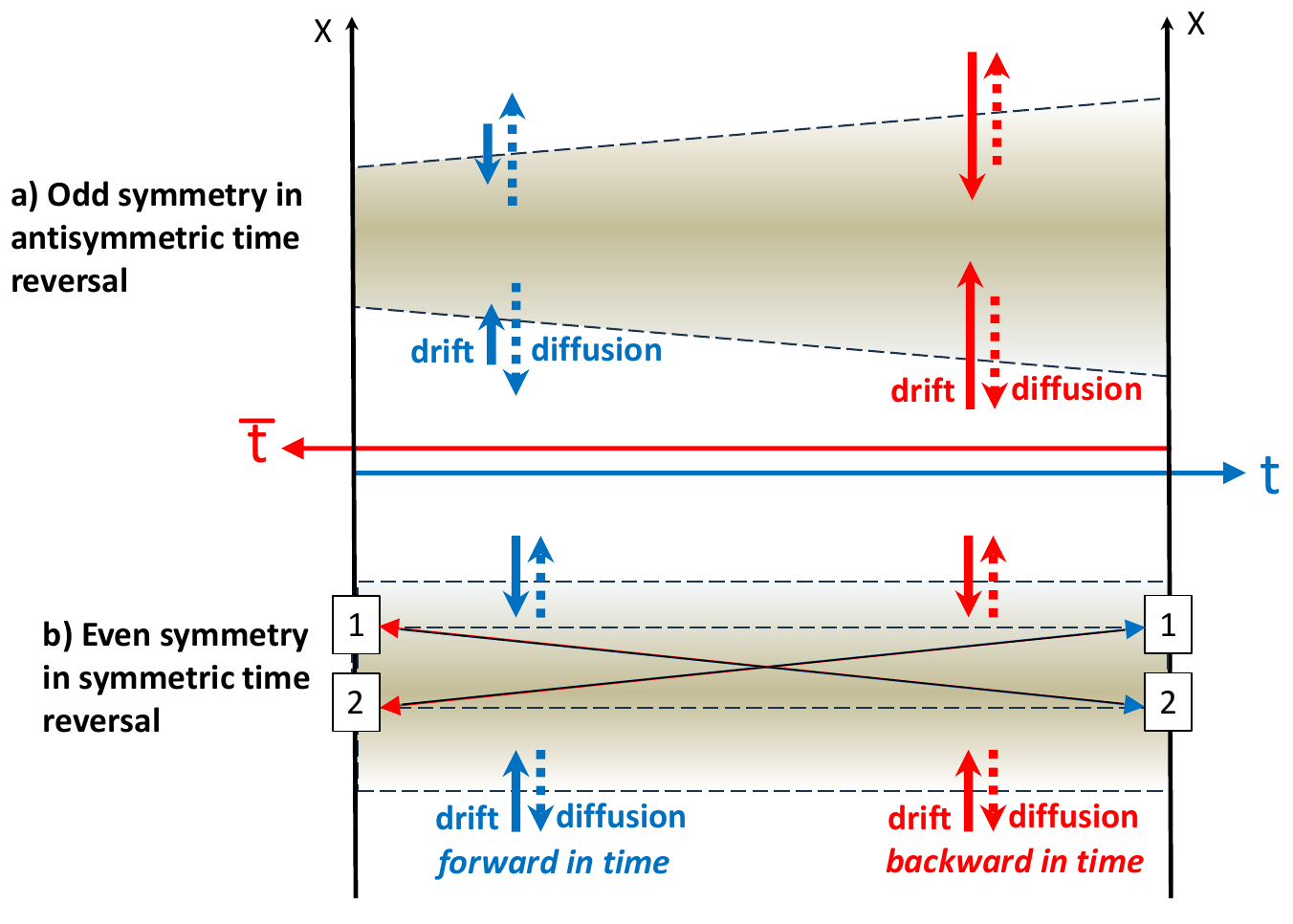}
\end{center}
\end{figure}

\subsection{Markov bridge model with odd temporal symmetry \label{Sec31}}

The true randomness is conventionally represented by a stochastic
\textit{Markov diffusion process} $\mathbf{x}_{t}=\mathbf{x}(t)$, governed by
a system of \textit{Stratanovich stochastic differential equations} (SDE) as
discussed in the Appendix. This process is considered within the time interval
$t\in\lbrack t_{1},t_{2}]$ subject to initial and final boundary conditions.

The probability distribution function is factorised $f(\mathbf{x}%
,t)=\varphi(\mathbf{x},t)\psi(\mathbf{x},t)/C.$ Under conditions specified
below,\ the functions $\varphi(\mathbf{x},t)$ and $\psi(\mathbf{x},t)$ satisfy
the corresponding direct-time and reverse-time Fokker-Planck (Kolmogorov
forward) equations
\begin{equation}
\frac{\partial\varphi}{\partial t}=-\frac{\partial a^{i}\varphi}{\partial
x^{i}}+\frac{\partial}{\partial x^{i}}\left(  B^{ij}\frac{\partial\varphi
}{\partial x^{j}}\right)  ,\ \ \text{with}\ \ \left(  \varphi\right)
_{t=t_{1}}=\delta(\mathbf{x}-\mathbf{x}_{1})\text{ }\label{FP-fi}%
\end{equation}%
\begin{equation}
\frac{\partial\psi}{\partial t}=-\frac{\partial a^{i}\psi}{\partial x^{i}%
}-\frac{\partial}{\partial x^{i}}\left(  B^{ij}\frac{\partial\psi}{\partial
x^{j}}\right)  ,\ \ \text{with}\ \ \left(  \psi\right)  _{t=t_{2}}%
=\delta(\mathbf{x}-\mathbf{x}_{2})\label{FP-psi}%
\end{equation}
These equations -- see Appendix A --- exhibit an antisymmetric correspondence
of the coefficients characterising the respective direct-time and reverse-time
stochastic processes. Here, $t_{1}\leq t\leq t_{2}$ and the parameters of the
model satisfy the following conditions ensuring odd symmetry
\begin{equation}
\frac{\partial a^{i}}{\partial x^{i}}=0,\ \ B^{ij}=\frac{b^{ik}b^{jk}}%
{2}=B^{ji},\ \ \frac{\partial b^{ik}}{\partial x^{i}}=0\label{FP-cond}%
\end{equation}
As stated in Proposition \ref{P3b}, these constrains ensure that the
formulation of the problem is exactly odd-symmetric: when time is reversed,
the model still specifies the same random process and no preference for the
direction of time is introduced by the model. Note that the diffusion matrix
$B^{ij}$ is symmetric and positive semidefinite and the drift coefficients
$a^{i}$ are Stratanovich drifts, not Ito drifts. Under odd-symmetric
conditions, the diffusion coefficients can be relatively small (and, as
discussed in Section \ref{sec2}, are physically expected to be small).
Constraints (\ref{FP-cond}) are both physical and mathematical. From physical
perspective, the constraint $\partial a^{i}/\partial x^{i}=0$ \ preserves the
phase volumes and entropy while $\partial b^{ik}/\partial x^{i}=0$ upholds the
well-mixed conditions. From mathematical perspective, constraints
(\ref{FP-cond}) ensure that the direct-time and reversed-time processes are
equivalent despite evolving in the opposite directions in time. This odd
symmetry of time reversal is not trivial: as time reverses, the diffusion and
diffusion-induced drift terms change sign while the sign of the principal
drift $a^{i}(\mathbf{x},t)$ remains the same in (\ref{FP-fi}) and
(\ref{FP-psi}). The conventional (Ito) formulation of SDE becomes
time-directional when $b^{ik}=$ $b^{ik}(\mathbf{x},t)$ depends on $\mathbf{x}$
and is not time-symmetric --- time reversal of stochastic trajectories
generally requires Stratanovich formulation of SDE. Therefore, these are
Stratanovich drifts $a^{i}$ and not Ito drifts $A^{i}$ that are deemed to
represent deterministic physical quantities.

The normalisation constant $C$ is determined by $C=\varphi(\mathbf{x}%
_{2},t_{2})=\psi(\mathbf{x}_{1},t_{1})$ so that the normalisation of $f$
\begin{equation}
\int fd\mathbf{x}=1,\ \ \ \ \ f=\frac{\phi\psi}{C}%
\end{equation}
is preserved by the model since equations (\ref{FP-fi}) and (\ref{FP-psi})
imply that equation for $f$ has a conservative form
\begin{equation}
\frac{\partial f}{\partial t}+\frac{\partial a^{i}f}{\partial x^{i}}%
=\frac{\partial}{\partial x^{i}}\left(  B^{ij}f\left(  \frac{\partial
\ln\varphi}{\partial x^{j}}-\frac{\partial\ln\psi}{\partial x^{j}}\right)
\right)  \label{FP-f}%
\end{equation}

The model allows us to introduce two entropies
\begin{equation}
S_{\varphi}(t)=\int\varphi\ln\left(  \frac{e}{\varphi}\right)
d\mathbf{x,\ \ }\ S_{\psi}(t)=\int\psi\ln\left(  \frac{e}{\psi}\right)
d\mathbf{x}%
\end{equation}
the direct-time entropy $S_{\varphi}$ and reverse-time entropy $S_{\psi}$. It
is easy to check that the direct-time entropy cannot decrease in time \
\begin{equation}
\frac{dS_{\varphi}}{dt}=-\int\frac{\partial\varphi}{\partial t}\ln\left(
\varphi\right)  d\mathbf{x=}\underset{=0}{\underbrace{\int\varphi
\frac{\partial a^{i}}{\partial x^{i}}d\mathbf{x}}}\mathbf{+}\underset{\Theta
_{\varphi}}{\underbrace{\int\frac{B^{ij}}{\varphi}\frac{\partial\varphi
}{\partial x^{i}}\frac{\partial\varphi}{\partial x^{j}}d\mathbf{x}}}%
=\Theta_{\varphi}\mathbf{\geq}0
\end{equation}
and reverse-time entropy cannot increase in time.%
\begin{equation}
\frac{dS_{\psi}}{dt}=-\int\frac{\partial\psi}{\partial t}\ln\left(
\psi\right)  d\mathbf{x=}\underset{=0}{\underbrace{\int\psi\frac{\partial
a^{i}}{\partial x^{i}}d\mathbf{x}}}\mathbf{-}\underset{\Theta_{\psi
}}{\underbrace{\int\frac{B^{ij}}{\psi}\frac{\partial\psi}{\partial x^{i}}%
\frac{\partial\psi}{\partial x^{j}}d\mathbf{x}}}=-\Theta_{\psi}\mathbf{\leq}0
\end{equation}
Conditions (\ref{div}) are essential for these conclusions. In evaluation of
the integrals by parts, we assume that that domain either infinite or, if it
is bounded, does not allow any probability fluxes through its boundaries, that
is the components of the drift and diffusion flux normal to the boundary
vanish. Assuming that $B^{ij}$ projection into the domain under consideration
is non-degenerate (and, therefore, positively defined), we note that
$dS_{\varphi}/dt=0$ can be only in a steady-state solution $\varphi
=\operatorname{const}.$ Similarly, \ $dS_{\psi}/dt=0$ is only when
$\psi=\operatorname{const}$. Hence, the equations admit a unique steady-state
solution that is constant, which can only be realised within domains of finite
measure due to the preserved normalisation of the distribution functions.

\textsl{Let us assume that such a domain corresponds to an isolated state of a
subsystem and lies on the associated manifold, whose dimension is determined
by the number of free degrees of freedom of the subsystem. This domain is
termed \textit{ergodically isolated} if neither dynamical
trajectories---assumed to satisfy the ergodicity conditions---nor diffusive
fluxes can cross its boundaries or escape the domain. Since the domain lies on
a manifold of reduced dimension, the condition of ergodic isolation implies
that there is no diffusion flux across the manifold. Consequently, the
diffusion matrix is degenerate, allowing diffusive fluxes along the manifold
but not across it. As noted above, the projection of }$B^{ij}$\textsl{ onto
the domain manifold is assumed to be non-degenerate. Given that homogeneous
Hamiltonian systems conserve energy, an ergodically isolated domain represents
an invariant manifold corresponding to a constant-energy surface. }

Since entropy is a monotonic and, in case of a finite volume, bounded function
of time, entropy monotonically converges to its maximal value $S(t)\rightarrow
S_{\text{max}}=\ln(V)$ in a finite measure (surface volume) $V,$ the
distributions $\varphi$ or $\psi$ should converge (as measured by the
corresponding \textit{Kullback--Leibler divergency}) to constant stationary
solutions forward in time for $\varphi$ and backward in time for $\psi$.

Under the conditions specified above, this behaviour of entropy can be
summarised by the following proposition:

\begin{proposition}
The odd-symmetric (antisymmetric) model has two entropies: direct-time entropy
$S_{\varphi}$ and reverse-time entropy $S_{\psi}$, which monotonically
increase forward in time and backward in time, respectively, until they
stabilise upon reaching steady-state solutions. These solutions are
necessarily constant and can only be achieved in finite-measure domains.
\end{proposition}

Identifying these distinct entropies $S_{\varphi}$ and $S_{\psi}$ might be
difficult as only the overall distribution $f$ $\sim\varphi\psi$ is directly
measurable. Hence we may wish to define a joint entropy
\begin{equation}
S_{f}(t)=\int f\ln\left(  \frac{e}{f}\right)  d\mathbf{x=}\underset{S_{f|\psi
}}{\underbrace{\int\frac{\psi}{C}\varphi\ln\left(  \frac{1}{\varphi}\right)
d\mathbf{x}}}\mathbf{+}\underset{S_{\psi|\varphi}}{\underbrace{\int%
\frac{\varphi}{C}\psi\ln\left(  \frac{1}{\psi}\right)  d\mathbf{x}}}%
+\ln\left(  eC\right)
\end{equation}
which represents a sum of the Kullback--Leibler divergencies $S_{f|\psi}$ and
$S_{f|\varphi}$, and may increase or decrease in time according to
\begin{align}
\frac{dS_{f}}{dt}  &  =-\int\frac{\partial f}{\partial t}\ln\left(  f\right)
d\mathbf{x=}\underset{=0}{\underbrace{\int f\frac{\partial a^{i}}{\partial
x^{i}}d\mathbf{x}}}+\int B^{ij}\frac{\partial f}{\partial x^{i}}\left(
\frac{\partial\ln\varphi}{\partial x^{j}}-\frac{\partial\ln\psi}{\partial
x^{j}}\right)  d\mathbf{x}\nonumber\\
&  =\underset{\Theta_{f|\psi}}{\underbrace{\int\frac{\psi}{C}\frac{B^{ij}%
}{\varphi}\frac{\partial\varphi}{\partial x^{i}}\frac{\partial\varphi
}{\partial x^{j}}d\mathbf{x}}}\mathbf{-}\underset{\Theta_{f|\varphi
}}{\underbrace{\int\frac{\varphi}{C}\frac{B^{ij}}{\psi}\frac{\partial\psi
}{\partial x^{j}}\frac{\partial\psi}{\partial x^{i}}d\mathbf{x}}}%
=\Theta_{f|\psi}-\Theta_{f|\varphi}%
\end{align}
The dissipation terms $\Theta_{f|\psi}\geq0$ and $\Theta_{f|\varphi}\geq0$ are
related to the corresponding dissipation terms $\Theta_{\varphi}$ and
$\Theta_{\psi}$ and vanish when the corresponding terms also vanish, so that
$\Theta_{f|\psi}=0$ corresponds to $\Theta_{\varphi}=0$ and $\Theta
_{f|\varphi}=0$ corresponds to $\Theta_{\psi}=0$.

The model (\ref{FP-fi})-(\ref{FP-f}) is consistent with dynamic constraint
(\ref{div}), which are inherited from fundamental mechanical models and
preserves the phase volume, as well as converges to uniform distributions in
the phase space (or to equidistribution between quantum states) when and if
thermodynamic equilibrium is achieved for a subsystem kept in an isolated
state in a domain of a finite measure. \textsl{Under ergodically isolated
conditions specified above and assumed valid in the rest of the paper, a subsystem should preserve its energy }%
$H$\textsl{, which corresponds exactly to microcanonical distribution in
statistical physics, implying equidistribution between all available states,
whose number is proportional to the phase volume.} The unusual feature of the
bridge model is presence of two entropies $S_{\varphi}$ and $S_{\psi},$ which
evolve in the opposite directions. This is a key feature of the model, which
is discussed in the next section.

\subsection{Markov bridge model with even temporal symmetry}

As detailed in Appendix B, the Kolmogorov's theory of even-symmetric temporal
reversibility of stationary Markov diffusion processes can be represented by
modified Fokker-Planck equations governing functions $\phi(\mathbf{x},t)$ and
$\psi(\mathbf{x},t)$
\begin{equation}
\eta\frac{\partial\phi}{\partial t}=\frac{\partial}{\partial x^{i}}\left(
\eta B^{ij}\frac{\partial\phi}{\partial x^{j}}\right)  ,\ \ \text{with}%
\ \ \left(  \phi\right)  _{t=t_{1}}=\delta(\mathbf{x}-\mathbf{x}_{1})
\label{even1}%
\end{equation}%
\begin{equation}
-\eta\frac{\partial\psi}{\partial t}=\frac{\partial}{\partial x^{i}}\left(
\eta B^{ij}\frac{\partial\psi}{\partial x^{j}}\right)  ,\ \ \text{with}%
\ \ \left(  \psi\right)  _{t=t_{2}}=\delta(\mathbf{x}-\mathbf{x}_{2})
\label{even2}%
\end{equation}
where $t_{1}\leq t\leq t_{2}$ and $\phi(\mathbf{x},t;\mathbf{x}_{1}%
,t_{1})=\eta\left(  \mathbf{x}_{1}\right)  \varphi(\mathbf{x},t;\mathbf{x}%
_{1},t_{1})/\eta\left(  \mathbf{x}\right)  $. Assuming $A^{i}=A^{i}%
(\mathbf{x})$ and $B^{ij}=B^{ij}(\mathbf{x})$, Kolmogorov
\cite{Kolmogoroff1937} established the following condition for reversibility
of the process
\begin{equation}
A^{i}\eta=\frac{\partial B^{ij}\eta}{\partial x^{j}}\text{ }\Longrightarrow
\tilde{A}^{i}=B^{ij}\frac{\partial\ln\eta}{\partial x^{j}} \label{KK-A}%
\end{equation}
where
\begin{equation}
\tilde{A}^{i}=A^{i}-\frac{\partial B^{ij}}{\partial x^{j}}=a^{i}-\frac{b^{ik}%
}{2}\frac{\partial b^{jk}}{\partial x^{j}},\ \ \ B^{ij}=\frac{b^{ik}b^{jk}}%
{2}=B^{ji},\ \ \ A^{i}=a^{i}+\frac{b^{jk}}{2}\frac{\partial b^{ik}}{\partial
x^{j}} \label{KK-AB}%
\end{equation}
and $\eta=\eta(\mathbf{x)}$ is a steady-state distribution that must exist. In
even-symmetric models, the diffusion $B^{ij}$ and drift $A^{i}$ coefficients
have similar magnitudes. The overall distribution $f$ is normalised with the
normalisation constant $C_{1}$ and governed by
\begin{equation}
f=\frac{\eta\phi\psi}{C_{1}},\ \ \ \frac{\partial f}{\partial t}%
=\frac{\partial}{\partial x^{i}}\left(  B^{ij}f\left(  \frac{\partial\ln\phi
}{\partial x^{j}}-\frac{\partial\ln\psi}{\partial x^{j}}\right)  \right)
\end{equation}

The direct-time and reverse-time entropies are defined by
\begin{equation}
S_{\phi}(t)=\int\eta\phi\ln\left(  \frac{e}{\phi}\right)  d\mathbf{x,\ \ }%
\ S_{\psi}(t)=\int\eta\psi\ln\left(  \frac{e}{\psi}\right)  d\mathbf{x}%
\end{equation}
and display monotonic behaviours
\begin{equation}
\frac{\partial S_{\phi}}{\partial t}=-\int\eta\frac{\partial\phi}{\partial
t}\ln\left(  \phi\right)  d\mathbf{x=}\int\frac{\eta}{\phi}B^{ij}%
\frac{\partial\phi}{\partial x^{i}}\frac{\partial\phi}{\partial x^{j}%
}d\mathbf{x=}\Theta_{\phi}\mathbf{\geq}0
\end{equation}%
\begin{equation}
\frac{\partial S_{\psi}}{\partial t}=-\int\eta\frac{\partial\psi}{\partial
t}\ln\left(  \psi\right)  d\mathbf{x}=-\int\frac{\eta}{\psi}B^{ij}%
\frac{\partial\psi}{\partial x^{i}}\frac{\partial\psi}{\partial x^{j}}%
=-\Theta_{\psi}\mathbf{\leq}0
\end{equation}
expected for proper entropies. The overall entropy
\begin{equation}
S_{f}(t)=\int f\ln\left(  \frac{e\eta}{f}\right)  d\mathbf{x=}%
\underset{S_{f|\psi}}{\underbrace{\int\eta\frac{\psi}{C_{1}}\phi\ln\left(
\frac{1}{\phi}\right)  d\mathbf{x}}}\mathbf{+}\underset{S_{f|\phi
}}{\underbrace{\int\eta\frac{\phi}{C_{1}}\psi\ln\left(  \frac{1}{\psi}\right)
d\mathbf{x}}}+\ln\left(  eC_{1}\right)
\end{equation}
changes consistently with partial entropies
\begin{align}
\frac{dS_{f}}{dt}  &  =-\int\frac{\partial f}{\partial t}\ln\left(  \frac
{f}{\eta}\right)  d\mathbf{x=}\int\frac{1}{\phi\psi}\frac{\partial(\phi\psi
)}{\partial x^{i}}\frac{\eta}{C_{1}}B^{ij}\left(  \psi\frac{\partial\phi
}{\partial x^{j}}-\phi\frac{\partial\psi}{\partial x^{j}}\right)
d\mathbf{x}\nonumber\\
&  =\underset{\Theta_{f|\psi}}{\underbrace{\int\frac{\eta}{C_{1}}\frac{\psi
}{\phi}B^{ij}\frac{\partial\phi}{\partial x^{i}}\frac{\partial\phi}{\partial
x^{j}}d\mathbf{x}}}\mathbf{-}\underset{\Theta_{f|\phi}}{\underbrace{\int%
\frac{\eta}{C_{1}}\frac{\phi}{\psi}B^{ij}\frac{\partial\psi}{\partial x^{j}%
}\frac{\partial\psi}{\partial x^{i}}d\mathbf{x}}}=\Theta_{f|\psi}%
-\Theta_{f|\phi}%
\end{align}
so that $\Theta_{f|\psi}\geq0$ and $\Theta_{f|\phi}\geq0$ are consistent with
the corresponding dissipations $\Theta_{\phi}\geq0$ and $\Theta_{\psi}\geq0$.
As in the previous subsection, monotonic properties of the entropies ensure
convergence to constant values of $\phi(\mathbf{x},t)$ forward in time and
$\psi(\mathbf{x},t)$ backward in time so that the overall distribution becomes
stationary $f(\mathbf{x},t)=\eta(\mathbf{x})$. Our analysis indicates that

\begin{proposition}
The even-symmetric model has two entropies: direct-time entropy $S_{\phi}$ and
reverse-time entropy $S_{\psi}$, which are introduced as the corresponding
Kullback--Leibler divergences from the $\mathbf{x}$-dependent steady-state
solution $\eta=\eta(\mathbf{x})$ monotonically increase forward in time
for$\ S_{\phi}$ and backward in time for $S_{\psi}$ until they stabilise upon
reaching the steady-state solution.
\end{proposition}

The first question is whether a model with true randomness can be both
even-symmetric and odd-symmetric simultaneously. This requires that
$\eta=\operatorname{const}$ and, therefore, $\tilde{A}^{i}$ and $a^{i}$ must
vanish in (\ref{KK-A}) and (\ref{KK-AB}) due to (\ref{FP-cond}). Hence,
possessing both symmetries is possible only in trivial cases, which are not of
interest here. Brownian bridge is an example of a trivial case that possesses
both symmetries, odd and even. When considering general laws of nature, the
requirements for both types of symmetry become opposing: potential character
of $\tilde{A}^{i}$ in (\ref{KK-A}) and solenoidal character of $a^{i}$ in
(\ref{FP-cond}). The both symmetries (of the odd and even types) can coexist
only in trivial cases or when applied to different degrees of freedom
associated with effectively independent subsystems. If a system is isolated,
then non-constant $\eta(\mathbf{x})$ is not consistent with the microcanonical
distribution and the observed laws of conventional thermodynamics. It seems,
however, that non-uniform equilibrium distributions associated with the
symmetric time reversal might be relevant to thermodynamic systems placed in
strong gravity (such as stars and black holes).

Our consideration can now be summarised by the following proposition:

\begin{proposition}
With the exception of trivial cases or applications to distinct autonomous
subsystems, the laws of nature involving true randomness cannot produce models
that are simultaneously symmetric and antisymmetric, i.e., that exhibit both
even and odd temporal symmetries. The antisymmetric model aligns well with
conventional thermodynamics and observed reality and must therefore be
dominant. The even-symmetric model might be relevant to autonomous systems
with strong gravitational effects.
\end{proposition}

\section{Time-symmetric models and their relevance to the observed universe}

While we consider two types of time-symmetric models in this section, the
system of equations (\ref{FP-fi})-(\ref{FP-psi}), which constitutes an
antisymmetric (or odd-symmetric) model, is of special interest here, as this
model is consistent with thermodynamic reality. In this model, the
deterministic component embodies the time-reversible dynamics of every minute
particle in nature implementing principal physical laws, while the random
component captures irreversible effects that drive changes in entropy.

This model exhibits temporal symmetry --- specifically of the odd or
antisymmetric type --- and, crucially, has time-symmetric temporal boundary
conditions. It also permits an explicit dependence on time, which may indicate
global expansion affecting the space-time metric or other fundamental changes
in the governing laws (assuming $t$ represents some kind of a synchronised time).

\subsection{The role of the final conditions in time-symmetric models}

In our approach, we consciously avoid the conventional pitfall of granting
unqualified primacy to initial conditions, which would inevitably imply the
presence of antecedent causality --- a concept that may hold relevance but
which we prefer not to accept as an unconditional postulate, leaving room for
it to emerge as a derived property within the framework.

The initial conditions at $t=t_{1}$ reflect the state of low entropy in the
early Universe, which is fundamental for understanding irreversible
thermodynamic behaviour driving the system towards equilibrium. This state may
or may not be related to the Big Bang theory; however, if it is, the key point
for us is that it was a low-entropy, well-ordered Big Bang. While the concept
of low-entropy beginnings has been widely accepted and discussed since
Boltzmann's time, the final conditions are rarely mentioned. This omission
corresponds to conventional thinking associated with antecedent causality.

In this work, we follow Schulman \cite{Schulman1973, Schulman1997} and impose
the final conditions at $t=t_{2}$ to mirror the initial conditions at
$t=t_{1}$, not because the final conditions are known or knowable, but to
avoid introducing time directionality and maintain symmetry in our
formulation. We therefore set the final conditions to be the same as, or
similar to, the initial conditions, assuming that the properties observed in
the present are not greatly affected by the precise details of these final
conditions.\ In simple terms, if we know nothing about the final conditions
and do not wish to discriminate between the directions of time a priori, we
must assume low-entropy final conditions to avoid building an a priori bias
into our theories of nature. Suggestions that the cosmological final state of
the Universe might be similar to its initial state have been repeatedly
discussed in publications
\cite{Gold1962,Hawking-Penrose,Schulman1997,Tamm2021}.

\subsection{The arrow of time and origin of causality.}

If both the initial and final conditions are set, the entropy is minimal at
the beginning, $t=t_{1}$, reaches its maximum midway at $(t_{1}+t_{2})/2$
(assuming full temporal symmetry) and decreases again at $t=t_{2}$ to its
initial value. Clearly, we are located somewhere at a time $t$ close to the
origin, meaning $t-t_{1}\ll t_{2}-t$, and we are influenced by the initial
conditions incomparably more strongly than by the final conditions. In the
representation $f=\varphi\psi/C$ (assuming odd symmetry), the function
$\varphi$ evolves actively forward in time according to the Fokker-Planck
equation and the fundamental laws of the Universe, thereby increasing the
entropy $S_{\varphi}$. In contrast, $\psi$ remains nearly constant everywhere,
and consequently, the reverse-time entropy $S_{\psi}$ is also nearly constant
so that the joint entropy $S_{f}\sim S_{\varphi}$ increases in time. We
observe non-equilibrium states induced by preselection effect of the initial
conditions, enabling us to make inferences about the early states of the
Universe. However, we do not observe any postselection associated with the
final conditions and have no certain knowledge about final conditions. The
potential for the entropy $S_{\varphi}$ to increase further suggests
substantial direct-time availability, whereas the reverse-time availability is
almost entirely absent.

Despite our time-symmetric approach, the conditions we describe are
practically indistinguishable from conventional interpretations. Suppose we
aim to solve the problem within the time interval $[t_{1}^{\prime}%
,t_{2}^{\prime}]$, where $t_{1}<t_{1}^{\prime}<t_{2}^{\prime}\ll t_{2}$. We
can set $\psi=\text{const}$ and $f\sim\varphi$ in (\ref{FP-f}), impose the
initial conditions at $t=t_{1}^{\prime}$, and solve the Fokker--Planck
equation in the form (\ref{FP-f}) forward in time from $t_{1}^{\prime}$ to
$t_{2}^{\prime}$. No final conditions are required. However, if we attempt to
specify the final conditions at $t=t_{2}^{\prime}$ instead of the initial
conditions, we face significant difficulties due to the ill-posed nature of
the problem. Consequently, we do not postulate antecedent causality but derive
its most crucial implication---our preference for setting initial conditions
rather than final conditions. Intuitively, we justify this principle by
observing that it is the past that influences the future, not vice versa.

The main point is now summarised in the proposition:

\begin{proposition}
The odd-symmetric interpretation of the laws governing the Universe, combined
with time-symmetric temporal boundary conditions without postulating
antecedent causality, gives rise to monotonic entropy growth, the arrow of
time, and the observable emergence of temporal causal order as long as we are
closer to the onset than to the end, i.e. $t_{1}<t\ll t_{2}.$
\end{proposition}

This proposition, which suggests the existence of a thermodynamic arrow of
time directed forward when $t_{1}<t\ll t_{2}$ and backward when $t_{1}\ll
t<t_{2},$ overlaps with the conclusions deduced from Schulman's model
\cite{Schulman1997}.

\subsection{Thermodynamic equilibrium in odd-symmetric models}

The thermodynamic behaviour is closely linked to the concept of equilibrium.
The reverse-time entropy and associated reverse-time thermodynamics are at or
near the global equilibrium state, whereas the direct-time entropy and
associated direct-time thermodynamics are not. Nevertheless, local or partial
equilibria are commonly achieved through direct-time evolution processes.

Following Penrose \cite{PenroseBook}, let us imagine the Universe as a maze of
constraints within a space of very high dimension, maintained by the
deterministic laws governing the Universe. The random component is represented
by diffusion. Direct-time evolution begins in a region of small volume and
progresses towards larger and larger volumes, increasing entropy $S_{\varphi}%
$. Within certain cells of the maze, equilibrium may be reached, implying a
uniform distribution within the cell but not between cells, meaning the system
remains far from global equilibrium.

More formally, a subsystem may become isolated within a time interval
$[t_{1}^{\prime},t_{2}^{\prime}]$, where $t_{1}<t_{1}^{\prime}<t_{2}^{\prime
}\ll t_{2}$, and thus preserves its total energy. For the degrees of freedom
associated with this subsystem, the Fokker--Planck equation (\ref{FP-fi})
evolves towards uniform distributions $f\sim\varphi=\text{const}$,
representing microcanonical equilibrium, while $\psi\sim\text{const
everywhere}$ within the interval. Such equilibrium may later be disturbed by
coupling the subsystem to others, continuing the direct-time evolution. The
reverse-time evolution remains negligible, as $\psi$ is very close to constant
throughout the accessible region.

\subsection{Distinctive features of the even-symmetric models}

While our previous consideration focusses on the antisymmetric model, it is
worthwhile to explore the implications of the second model with even temporal
symmetry. In many respects, the symmetric model behaves similarly to the
antisymmetric model, but its equilibrium becomes a joint property of
direct-time and reverse-time evolutions (rather than their more autonomous
properties, as in the antisymmetric model) implying a stronger conceptual
coupling of the direct-time and reverse-time submodels. This equilibrium,
$f=\eta(\mathbf{x})$, is significantly non-uniform, while a non-equilibrium
state is expressed as the product $f\sim\eta\phi\psi$ involving two
multiplicative functions: $\phi$, evolving forward in time, and $\psi$,
evolving backward in time.

Considering that distributions in phase spaces of very large dimension are
usually sharp, only regions where $\eta(x)\approx\eta_{\max}$ are
statistically significant, effectively reducing the entire space to the
vicinity of a single point or a relatively small set of points where
$\eta(x_{\max})=\eta_{\max}$. Despite this highly localised state, where
$\phi\sim\psi\sim\text{const}$ and $f\sim\eta(x)$, the entropy $S_{f}$ reaches
its maximum. This behaviour, although unusual from a conventional
thermodynamic perspective, is indeed observed in nature. For example, black
holes exhibit a very limited set of defining parameters but have the highest
possible entropy. Here, we suggest only that the even type of temporal
symmetry may be relevant to black holes and, of course, not that equations
(\ref{even1})-(\ref{even2}) specifically govern them.

Another key feature of the symmetric model is strong dissipation. Although
random effects are often assumed to be small, allowing deterministic equations
to remain reasonably accurate, there are mechanisms of amplification of the
diffusion effects due to dynamic mixing. In the symmetric model, there is
additional amplification mechanism associated with strong convergence of the
drift producing sharp gradients and ensuring that deterministic and random
terms remain of the same order. This intensifies dissipation, and for a given
characteristic energy $E$, results in high entropy changes $\Delta S$,
implying very low temperatures, $T\sim E/\Delta S $. Once again, such unusual
behaviour is conventionally associated with black holes.

\section{The effects associated with the reverse-time part of the model}

Despite maintaining temporal symmetry, the behaviour discussed in the previous
section is indistinguishable from conventional entropy-increasing evolution,
while the final conditions remain hidden. This section aims to explore
potential effects associated with the model's reverse-time dynamics. To
observe such effects, one might consider investigating the remote future
$t_{1}\ll t<t_{2},$\ \ This, however, is of limited interest, as thermodynamic
behaviour in the distant future would closely mirror the present, with the
arrow of time simply reversed.

\textsl{At this point, it is important to clarify that the influence of the
initial boundary conditions on the random terms generally remains indefinite.
When constrained in the past, these random components increasingly propagate
uncertainty into the future, a behaviour that corresponds to the direct
parabolic terms in the Fokker--Planck equation. While the relative magnitude
of these terms may diminish, their fundamental nature remains unchanged,
consistently reflecting direct-time physics. These direct parabolic terms do
not gradually transform into reverse-parabolic terms; rather, the
direct-parabolic and reverse-parabolic components coexist, competing with one
another. A similar consideration applies to the final conditions --- their
influence does not entirely vanish even during early stages. Although the
relative magnitude of these effects may be small, they could, in principle, be
detected, albeit with extreme --- if not insurmountable --- difficulty in
practice.}

Accordingly, the principal conceptual question is how the reverse-time
component of the antisymmetric model could manifest in present conditions, no
matter how subtle or insignificant these effects might be. Schulman
\cite{Schulman1997} suggested that both direct-time and reverse-time effects
may coexist. Tamm \cite{Tamm2021} proposed that the future state of the
Universe could influence its current rate of expansion --- this is a related
but distinct issue concerning the deterministic dynamics of the Universe. In
contrast, our analysis focuses on detecting subtle effects associated with
randomness and linked to the post-selection of the final conditions.
Specifically, we seek to understand how such effects might appear to an
observer when direct-time evolution remains dominant and preserves a strong
conventional arrow of time. There are two cases of interest: emergence of a
small global gradient of function $\psi$ and a local variation of $\psi$ in a
tiny isolated system, while $\psi$ remains (nearly) constant everywhere.

\subsection{Detecting post-selection}

Let us assume that, although $\psi$ is close to a constant, there exists a
small global gradient of $\psi$, while $\varphi$ evolves intensively forward
in time. First of all, the presence of a global gradient in $\psi$ affects
only random processes and has no influence on deterministic dynamics.
Therefore, our fundamental mechanical laws remain unchanged. The existence of
a gradient in $\psi$ implies that the post-selection of random trajectories by
the final temporal boundary conditions introduces some preferences or biases.
For example, post-selection could, at least in principle, create a slight
preference for heads over tails. If so, tossing a perfectly fair coin many
times would reveal a tiny bias towards heads---provided that we could detect
even the smallest bias, which is doubtful since a physical coin is unlikely to
behave ideally. Another issue is why the final conditions would favour heads.
If the toss outcomes---heads or tails---form a direction in phase space that
is perpendicular to the gradient of $\psi$, then the final conditions imposed
on the Universe would have no effect on the coin's outcome, which,
practically, is the most likely scenario.

A more precise set of experiments can be conducted by measuring quantum spin,
assuming that, by default, spin up $\left\vert \uparrow\right\rangle $ and
spin down $\left\vert \downarrow\right\rangle $ are equally probable outcomes.
Theoretically, post-selection can introduce certain biases. Our particle may
inadvertently be entangled with another particle elsewhere in the universe.
This distant particle could be influenced by the final conditions, for
instance, through the post-selection of a particular magnetic field in the
final stages of the universe. It goes without saying that detecting such an
effect remains purely a hypothetical possibility.

Some may interpret the reverse-time evolution of $\psi$ as introducing
retrocausality. Even if we accept this perspective, conducting experiments to
detect retrocausality is problematic, as the concept of causality, while
useful, is rather vague and largely based on intuition. Moreover, the
conventional "flow of causality", associated with the evolution of $\varphi$,
would overwhelm any other effects.

Consider a system that remains isolated during the time interval
$[t_{1}^{\prime},t_{2}^{\prime}].$ Initially, at $t=t_{1}^{\prime}$, it is
generally not in equilibrium (due to preselection by the low-entropy initial
conditions imposed on the Universe). However, after some time $t=t_{1}%
^{\prime}+\Delta t$, the equilibrium distribution $\varphi
=\operatorname{const}$ is established. Similarly, postselection could be
detected only after $t=t_{2}^{\prime}-\Delta t$, as $\psi$ would reach
equilibrium $\psi=\operatorname{const}$ backwards in time when $t<t_{2}%
^{\prime}-\Delta t$. In practice, an experimentalist would observe an increase
in thermodynamic fluctuations just before opening the system to the outside
world. $\ $

Our Universe has existed for 13 billion years and is still far from
equilibrium. This suggests that, if final conditions do exist, they must lie
far beyond 13 billion years into the future. It is therefore highly likely
that these conditions have no appreciable postselection effect at present.
Even if they exerted some minor influence, detecting postselection would
remain challenging.

\subsection{Observing localised antisystems}

The direct-time and reverse-time components seem to be linear and independent,
coupled only through the joint probability distribution $f=\varphi\psi/C$.
However, this apparent independence is misleading and typically arises when
Liouville-type equations are applied in spaces of extremely high dimension. As
soon as we attempt to reduce the number of variables to a practically
justifiable level, the subsystems exhibit strong nonlinear interactions, as
illustrated in the following example.

Consider a small system that is completely isolated---this system reaches
internal equilibrium $\varphi_{s}=\operatorname{const}$ and can, in principle,
exist in this state indefinitely, even if it is not in equilibrium with the
universe, i.e. $\varphi_{s}\neq\varphi_{u}$. Similarly, an isolated
\textit{antisystem} with constant $\psi_{a}\neq\psi_{u}$ could, at least in
principle, exist in our time. While such systems and antisystems can exist
theoretically, preserving their properties in a hostile environment over long
durations seems problematic. However, another possibility arises from the
stochastic nature of randomness, reflected in the presence of fluctuations.
Therefore, tiny systems and antisystems that are not in equilibrium with their
environments may occasionally appear (and disappear) due to fluctuations,
provided that such appearances comply with the deterministic laws of the Universe.

What properties would such a tiny antisystem exhibit if it appeared around us?
Our world is characterised by varying $\varphi=\varphi(x,t)$ and nearly
constant $\psi=\psi_{u}=\operatorname{const}$, whereas the antisystem has
$\psi_{a}>\psi_{u}$. Since the antisystem is small, the global laws of
thermodynamics should remain valid as long as its behaviour can be described
in conventional thermodynamic terms. On the surface, this seems problematic
since $S_{\psi}$ tends to decrease in time, behaving unconventionally.
However, instead of the symmetric entropy $S_{f}$, which behaves
inconsistently within our thermodynamic framework, we can define an effective
or antisymmetric entropy:
\begin{equation}
S_{\text{eff}}=S_{\varphi}-S_{\psi} \label{SSS}%
\end{equation}
which remains monotonic in time and cannot decrease, i.e., $dS_{\text{eff}%
}/dt\geq0$. With this definition of entropy, the conventional laws of
thermodynamics remain intact.

Let us consider the effect of energy exchange $\delta Q$ from a small
antisystem with $\psi_{a}>\psi_{u}$ and $\varphi_{a}=\varphi_{s}$ to a system
with $\varphi_{s}$ and $\psi_{s}=\psi_{u}$ so that the system and antisystem
energies become after the exchange $E_{s}+\delta Q$ and $E_{a}-\delta Q$
respectively. Considering autonomous, intrinsic properties of these systems,
the entropy variation due to energy change can be expressed as:%
\begin{equation}
dS_{s}=\frac{dE_{s}}{T_{s}},\quad dS_{a}=\frac{dE_{a}}{T_{a}}%
\end{equation}
where the energies, entropies, and temperatures are the corresponding
intrinsic properties of the system and antisystem. Since $dE_{s}=\delta Q$ and
$dE_{a}=-\delta Q$, we obtain:
\begin{equation}
dS_{\text{eff}}=dS_{s}-dS_{a}=\left(  \frac{1}{T_{s}}+\frac{1}{T_{a}}\right)
\delta Q\geq0.
\end{equation}

Unlike in conventional thermodynamic interactions, where $dS=\left(
T_{s1}^{-1}-T_{s2}^{-1}\right)  \delta Q$, the sign of $dS$ depends on the
temperatures of the interacting systems s1 and s2. In contrast,
$dS_{\text{eff}}$ is always positive, strongly favouring energy transfer from
the antisystem to the system. This process does not terminate even if the
intrinsic temperature of the antisystem becomes low, continuing until the
antisystem loses all its energy. Note that a system cannot simultaneously
possess both direct-time and reverse-time availabilities as this would be
thermodynamically unstable due to immediate contact of the subsystems. While
specialised antisymmetric version of thermodynamics and kinetics has been
developed for the interactions between thermodynamic systems and antisystems
\cite{KM-Entropy2014,SciRep2016, Ent2017, Klimenko2021}, our objective here is
to highlight this unusual behaviour and determine whether it may manifest in
the known Universe. The complete transfer of energy into the environment
strongly resembles properties of antimatter in its interactions with matter.
We may indeed have antisystems present in our world in the form of
antiparticles which, according to Feynman, can be deemed to "travel" to us
from the future. The theory introducing an antisymmetric extension of
thermodynamics and kinetics from systems to antisystems is experimentally
testable, even at the present level of technologyby
\cite{KM-Entropy2014,SciRep2016, Ent2017, Klimenko2021}.

\textsl{The analysis of this section is summarised by the following
proposition: }

\begin{proposition}
\textsl{The direct-time and reverse-time random effects, respectively
characterised by }$\varphi$ \textsl{and} $\psi,$\textsl{ tend to persist
indefinitely and conceptually coexist without blending or converting into one
another, while remaining antagonistic from a thermodynamic perspective. The
entities governed by these direct-time and reverse-time effects are
respectively termed thermodynamic systems and antisystems. The practical
detection of reverse-time effects remains problematic under present
conditions, which are dominated by the direct arrow of time --- unless compact
thermodynamic antisystems, predicted by the bridge equations, actually exist in our world, in which case
it should be possible to examine such properties experimentally. The
fundamental question of whether antimatter possesses the properties of
thermodynamic antisystems remains open. }
\end{proposition}

\section{Conclusions}

Our present analysis is based on the following assumptions:

\begin{enumerate}
\item Most of the \textit{fundamental physical laws} governing nature
(encompassing classical, relativistic, and quantum mechanics) are
\textit{deterministic} (implying quantum unitary determinism), time-reversible
(i.e. they preserve entropy) and are close to being time-symmetric.

\item \textit{Randomness} is inherently present in nature, as reflected in
thermodynamics, statistical physics, and in some irreversible quantum effects.

\item The Universe is subject to \textit{low-entropy initial conditions} in
the distant past.

\item The Universe is also subject to some \textit{low-entropy final
conditions} in an even more distant future (i.e. the \textit{Schulman
condition}).
\end{enumerate}

Some of these assumptions --- specifically assumptions 1 and 3 --- are more
conventional. We explore the implications of adding assumptions 2 and 4 ---
omnipresent randomness and the Schulman condition --- which remove the
inherent asymmetry of time found in most conventional theories and allow us to
demonstrate the \textit{arrow of time} without postulating antecedent
causality. The ubiquitous presence of randomness is responsible for
time-directional effects and inevitably leads to \textit{Fokker--Planck}-type
equations in high-dimensional spaces. Further analysis, based on temporal
symmetry, results in two principal types of models:

\begin{description}
\item[A. Antisymmetric (odd-symmetric) models,] which characterise
conventional thermodynamic behaviour

\item[B. Symmetric (even-symmetric) models,] which may be associated with
thermodynamic systems in strong gravitational fields.
\end{description}

Our analysis explains the emergence of the world as we know it, encompassing
conventional thermodynamics, the arrow of time, and a form of antecedent
causality---all of which are deduced rather than postulated. This behaviour,
associated with the relatively early stages of the Universe's evolution,
arises from mathematical properties of \textit{Markov bridges} and does not
rely on common assumptions that impose a preferred direction of time.

Additionally, we explore the effects of reverse-time dynamics associated with
final conditions, \textsl{which are presumably far more remote than the
initial conditions,} and find that such effects would be difficult to detect
in the real world---except through the emergence of antisystems whose
properties closely resemble those of antimatter. Whether antimatter possesses
the properties of thermodynamic antisystems is experimentally testable.


\appendix


\renewcommand{\theequation}{A\arabic{equation}}
\setcounter{equation}{0} \setcounter{section}{0}

\section{Notes on Bridge Conditions in Markov Models\label{appA}}

This Appendix overviews mathematical properties of \textit{{Markov diffusion
bridges}} that are used in the main text establishing conditions for invariant
time reversal.

\subsection{Stochastic Equations Forward and Backward in Time}

Consider a family of stochastic \textit{{Markov diffusion processes}}
$\mathbf{x}_{t}$ that has the same semigroup generator coefficients
$a^{i}(\mathbf{x}_{t},t)$ and $b^{ik}(\mathbf{x}_{t},t),$ and is governed by
the following system of \textit{{Ito stochastic differential equations}} (SDEs)
\begin{equation}
dx_{t}^{i}=A^{i}(\mathbf{x}_{t},t)dt+b^{ik}(\mathbf{x}_{t},t)dw_{t}^{k}
\label{sde-If}%
\end{equation}
{The} default Einstein summation convention applies here and throughout the
Paper, while $w_{t}^{k}$ denotes independent \textit{{Wiener processes}}. In
addition to the process (\ref{sde-If}) defined forward in time, we can
introduce the reversed time $\bar{t}=-t$\ and run another process
$\mathbf{\bar{x}}_{\bar{t}}$ forward in time $\bar{t}$ and backward in time
$t$, as specified by the stochastic differential equation
\begin{equation}
d\bar{x}_{\bar{t}}^{i}=\bar{A}^{i}(\mathbf{\bar{x}}_{\bar{t}},\bar{t})d\bar
{t}+\bar{b}^{ik}(\mathbf{\bar{x}}_{\bar{t}},\bar{t})d\bar{w}_{\bar{t}}^{k}
\label{sde-Ib}%
\end{equation}
{The} drift and diffusion coefficients $\bar{A}^{i}$ and $\bar{b}^{ik}$\ are
generally different from $A^{i}$ and $b^{ik}$. While we are interested in
cases and conditions under which the processes $\mathbf{\bar{x}}_{\bar{t}%
}(-t)$ and $\mathbf{x}_{t}(t)$ are equivalent, we do not imply yet at this
stage any equivalence or similarity of $\mathbf{\bar{x}}_{\bar{t}}$ and
$\mathbf{x}_{t}$ and treat these processes in (\ref{sde-If}) and
(\ref{sde-Ib}) as being different. As demonstrated by Anderson~\cite{Anderson1982}, $\bar{w}_{\bar{t}}^{k}$ is related to $w_{t}^{k}$\ but
with some adjustments to preserve the retrocasual structure of the
reverse-time stochastic integrals and ensure that $\mathbf{\bar{x}}_{\bar{t}%
}(-t)=\mathbf{x}_{t}(t)$ in a strong sense. Here, we just assume that $\bar
{w}_{\bar{t}}^{k}$ is another equivalent Wiener process.

It should be noted that Ito calculus is time-directional enforcing
independence of $\mathbf{x}_{t}(t)$ of $\mathbf{w}_{t}(t^{\prime})$ when
$t^{\prime}>t$. This understanding is implemented in the numerical definition
of the product $b^{ik}(\mathbf{x}_{t},t)dw_{t}^{k}$ in the corresponding
stochastic integral, where $b^{ik}(\mathbf{x}_{t},t)$ is evaluated using the
value of $\mathbf{x}_{t}$ at the beginning of each time step. A more general
approach can be formally expressed as $b^{ik}(\mathbf{x}_{t},t)_{\gamma}%
dw_{t}^{k}$, implying that \ $b^{ik}$ is evaluated as $b^{ik}(\mathbf{x}%
_{t}+\gamma d\mathbf{x}_{t},t)dw_{t}^{k},$ where $0\leq\gamma\leq1$ and
$\gamma=0$ corresponds to the Ito definition, $\gamma=1/2$ to the Stratanovich
definition, and $\gamma=1$ to the anti-Ito definition of the stochastic
integral and the corresponding differential equations. Note that, when time is
reversed, the Ito definition is converted into the anti-Ito definition, 
rather than yielding the Ito definition evaluated backward in time, that is
$\bar{\gamma}=1-\gamma$. Only the \textit{Stratanovich formulation of SDEs},
which has $\bar{\gamma}=\gamma=1/2$ and is conventionally denoted by
$b^{ik}(\mathbf{x}_{t},t)\circ dw_{t}^{k},$ is time-symmetric, corresponding
to the same Stratanovich formulation when evaluated backward in time. Note
that while Stratanovich calculus removes some temporal asymmetry of the Ito
calculus, it is still conceptually related to Ito calculus and, similar to Ito
calculus, has some inherent directionality of time built-in into its
foundation. The Stratanovich SDE that corresponds to (\ref{sde-If}) is
specified by the following equation with the new drift $a^{i}$ adjusted by
removing diffusion-generated drift component from $A^{i}$%
\begin{equation}
dx_{t}^{i}=a^{i}(\mathbf{x}_{t},t)dt+b^{ik}(\mathbf{x}_{t},t)\circ dw_{t}%
^{k},\ \ \ \ \ A^{i}(\mathbf{x}_{t},t)=a^{i}(\mathbf{x}_{t},t)+\frac{1}%
{2}b^{jk}(\mathbf{x}_{t},t)\nabla_{j}b^{ik}(\mathbf{x}_{t},t) \label{sde-f}%
\end{equation}
{Here}, and further in this appendix, $\nabla_{j}$ denotes the partial
derivative $\partial/\partial x^{j}.$ The relation between Ito drift $A^{i}$
and Stratanovich drift $a^{i}$ can be easily obtained by expanding $b^{ik}$
into a series and substituting $\gamma=1/2$ and $dw_{t}^{l}dw_{t}^{k}%
=\delta^{lk}dt$. The Stratanovich SDE for the time-reversed process
(\ref{sde-Ib}) is given by a similar equation
\begin{equation}
d\bar{x}_{\bar{t}}^{i}=\bar{a}^{i}(\mathbf{\bar{x}}_{\bar{t}},\bar{t})d\bar
{t}+\bar{b}^{ik}(\mathbf{\bar{x}}_{\bar{t}},\bar{t})\circ d\ \bar{w}_{\bar{t}%
}^{k},\ \ \ \ \bar{A}^{i}(\mathbf{\bar{x}}_{\bar{t}},\bar{t})=\bar{a}%
^{i}(\mathbf{\bar{x}}_{\bar{t}},\bar{t})+\frac{1}{2}\bar{b}^{jk}%
(\mathbf{\bar{x}}_{\bar{t}},\bar{t})\nabla_{j}\bar{b}^{ik}(\mathbf{\bar{x}%
}_{\bar{t}},\bar{t}) \label{sde-b0}%
\end{equation}
where $d\bar{w}_{\bar{t}}^{l}d\bar{w}_{\bar{t}}^{k}=\delta^{lk}d\bar{t}$ and
$\bar{\gamma}=1-\gamma=1/2$ are used. Note that these are the Stratanovich
drift coefficients $\bar{a}^{i}$ and $a^{i}$ that are treated here as physical
quantities, while the Ito drift coefficients $\bar{A}^{i}$ and $A^{i}$ involve
mathematical corrections merely reflecting temporal asymmetry of the Ito
formulation. As one can see below, the physical drift term and the
diffusion-induced drift adjustment change differently under time reversal.

It is also worthwhile to consider the process $\mathbf{\bar{x}}_{\bar{t}}$
defined by (\ref{sde-b0}) as being parametrised by direct time $t=-\bar{t}$,
that is $\mathbf{\bar{x}}_{t}(t)=\mathbf{\bar{x}}_{\bar{t}}(-t)$. Note that
$\mathbf{\bar{x}}_{t}$ is generally different from $\mathbf{x}_{t}$. In line
with antisymmetric time reversal (\ref{A12}), we can assume
\begin{equation}
\bar{a}^{i}(\mathbf{x},-t)=-a^{i}(\mathbf{x},t),\text{ \ \ \ \ }\bar{b}%
^{ij}(\mathbf{x},-t)=-b^{ij}(\mathbf{x},t) \label{aabb}%
\end{equation}
use $dt=-d\bar{t}$ and $\bar{d}w_{t}^{k}=-\bar{d}w_{\bar{t}}^{k}$ to obtain
\begin{equation}
d\bar{x}_{t}^{i}=a^{i}(\mathbf{\bar{x}}_{t},t)dt+b^{ik}(\mathbf{\bar{x}}%
_{t},t)\circ d\bar{w}_{t}^{k} \label{sde-b}%
\end{equation}
{This} is still reverse-time process $\mathbf{\bar{x}}_{t}$, but parametrised by
$t$ instead of $\bar{t},$ while the processes $\mathbf{x}_{t}$\ and
$\mathbf{\bar{x}}_{t}$\ may or may not be equivalent. The apparent similarity
of Equations (\ref{sde-b}) and (\ref{sde-f}) is deceiving: note that
$dw_{t}^{l}dw_{t}^{k}=\delta^{lk}dt$ in (\ref{sde-f}) but $d\bar{w}_{t}%
^{l}d\bar{w}_{t}^{k}=(-d\bar{w}_{\bar{t}}^{l})(-d\bar{w}_{\bar{t}}^{k}%
)=\delta^{lk}d\bar{t}=-\delta^{lk}dt.$\ This is a principal change associated
with time reversal of diffusion, which cannot be removed by a mere replacement
of variables.

Note that the antisymmetric assumption $\bar{a}^{i}=-a^{i}$ is consistently
applied here to the Stratanovich drifts while the Ito drifts $\bar{A}^{i}%
\neq-A^{i}$ are generally different, assuming $\bar{A}^{i}=-A^{i}$ would
constitute a different physical assumption that would result in a more
restrictive understanding and equations.

\subsection{Transitional Probabilities and Markov Properties}

The \textit{{Markov property}} is fundamental to the stochastic equations and is
conventionally interpreted in time-directional manner: the future does not
depend on the past when the present is given. Note that this also means that
the past is independent of the future when the present is given.\ Assuming
$t_{1}<t<t_{2}$ we can write for conditional probabilities of $\mathbf{x}%
_{t}$
\begin{equation}
\mathbb{P}\left(  \mathbf{x}_{2},t_{2}|\mathbf{x}_{1},t_{1};\mathbf{x}%
,t\right)  =\mathbb{P}\left(  \mathbf{x}_{2},t_{2}|\mathbf{x},t\right)
\label{M1}%
\end{equation}
or equivalently
\begin{equation}
\mathbb{P}\left(  \mathbf{x}_{1},t_{1}|\mathbf{x},t;\mathbf{x}_{2}%
,t_{2}\right)  =\mathbb{P}\left(  \mathbf{x}_{1},t_{1}|\mathbf{x},t\right)  .
\label{M2}%
\end{equation}
{Here}, ($\mathbf{x}_{k},t_{k}$) denotes condition $\mathbf{x}_{t}%
(t_{k})=\mathbf{x}_{k}$ and $\mathbb{P}\left(  \text{A}|\text{B}\right)  $
represents the conditional (or transitional in the Markov context) probability of
event A given event B. Note that independence of the future from the past also
means that the past does not depend on the future when the present is given.
Hence, both Equations (\ref{M1}) and (\ref{M2}) are valid for both
processes.\ In time-symmetric interpretations, the Markov property resembles
properties of Gibbs fields~\cite{Georgii2011} on linear topology: this
property implies that for any $t_{0}<t_{1}<t<t_{2}<t_{3}$ by
\begin{equation}
\mathbb{P}\left(  \mathbf{x},t|\mathbf{x}_{0},t_{0};\mathbf{x}_{1}%
,t_{1};\mathbf{x}_{2},t_{2};\mathbf{x}_{3},t_{3}\right)  =\mathbb{P}\left(
\mathbf{x},t|\mathbf{x}_{1},t_{1};\mathbf{x}_{2},t_{2}\right)  \label{M12}%
\end{equation}
{While} the Markov property is objectively time-symmetric, all principal
formulations of stochastic calculus---the Ito, Stratonovich, and Skorokhod
versions---incorporate at least some time-directionality, albeit to different extents.

The application of conditional version of the probabilistic rules to the
process $\mathbf{x}_{t}$ within, assuming that the times are ordered as
$t_{1}<t<t_{2}$, gives
\begin{equation}
\mathbb{P}\left(  \mathbf{x},t|\mathbf{x}_{1},t_{1};\mathbf{x}_{2}%
,t_{2}\right)  =\mathbb{P}\left(  \mathbf{x}_{2},t_{2}|\mathbf{x}_{1}%
,t_{1};\mathbf{x},t\right)  \frac{\mathbb{P}\left(  \mathbf{x},t|\mathbf{x}%
_{1},t_{1}\right)  }{\mathbb{P}\left(  \mathbf{x}_{2},t_{2}|\mathbf{x}%
_{1},t_{1}\right)  }=\frac{\mathbb{P}\left(  \mathbf{x}_{2},t_{2}%
|\mathbf{x},t\right)  \mathbb{P}\left(  \mathbf{x},t|\mathbf{x}_{1}%
,t_{1}\right)  }{\mathbb{P}\left(  \mathbf{x}_{2},t_{2}|\mathbf{x}_{1}%
,t_{1}\right)  } \label{PP-f}%
\end{equation}
which uses Markov property (\ref{M1}) and reflects conventional probability
manipulations according to Bayes theorem $\mathbb{P}\left(  \text{A}%
|\text{B}\right)  =\mathbb{P}\left(  \text{B}|\text{A}\right)  \mathbb{P}%
\left(  \text{A}\right)  /\mathbb{P}\left(  \text{B}\right)  $. \ We also need
a similar expression for the reverse-time process $\mathbf{\bar{x}}_{t}$: note
that the Markov properties (\ref{M1})--(\ref{M12}) are equivalently valid for
$\mathbf{\bar{x}}_{t}.$ Equation (\ref{PP-f}) then becomes
\begin{equation}
\mathbb{P}\left(  \mathbf{\bar{x}},t|\mathbf{\bar{x}}_{1},t_{1};\mathbf{\bar
{x}}_{2},t_{2}\right)  =\frac{\mathbb{P}\left(  \mathbf{\bar{x}}_{1}%
,t_{1}|\mathbf{\bar{x}},t\right)  \mathbb{P}\left(  \mathbf{\bar{x}%
},t|\mathbf{\bar{x}}_{2},t_{2}\right)  }{\mathbb{P}\left(  \mathbf{\bar{x}%
}_{1},t_{1}|\mathbf{\bar{x}}_{2},t_{2}\right)  } \label{PP-b}%
\end{equation}
where Markov property (\ref{M2}) is applied to the process $\mathbf{\bar{x}%
}_{t}$ and ($\mathbf{\bar{x}}_{k},t_{k}$) denotes the condition
$\mathbf{\bar{x}}_{t}(t_{k})=\mathbf{x}_{k}$.

\subsection{Kolmogorov Equations for Transitional Probabilities
\label{sec A-Ke}}

The transitional probabilities for a Markov diffusion process depend only on
diffusion and drift coefficients and on the conditions that explicitly define
these probabilities but not on any other initial or final conditions that may
constrain the overall distribution. The transitional probabilities
$\varphi\left(  \mathbf{x},t;\mathbf{x}_{1},t_{1}\right)  =\mathbb{P}\left(
\mathbf{x},t|\mathbf{x}_{1},t_{1}\right)  $ and $\psi\left(  \mathbf{x}%
,t;\mathbf{x}_{2},t_{2}\right)  =\mathbb{P}\left(  \mathbf{x}_{2}%
,t_{2}|\mathbf{x},t\right)  $ of the direct-time Markov diffusion process
$\mathbf{x}_{t}$ defined by Ito SDE (\ref{sde-If}), respectively, satisfy the
following \textit{{Kolmogorov forward and backward equations}}
\cite{Kolmogoroff1937}
\begin{equation}
\frac{\partial\varphi}{\partial t}=-\nabla_{i}\left(  A^{i}\varphi\right)
+\nabla_{i}\nabla_{j}\left(  B^{ij}\varphi\right)  ,\ \ \text{with}\ \ \left(
\varphi\right)  _{t=t_{1}}=\delta(\mathbf{x}-\mathbf{x}_{1}) \label{fi1}%
\end{equation}%
\begin{equation}
-\frac{\partial\psi}{\partial t}=A^{i}\nabla_{i}\psi+B^{ij}\nabla_{i}%
\nabla_{j}\psi,\ \ \text{with}\ \ \left(  \psi\right)  _{t=t_{2}}%
=\delta(\mathbf{x}-\mathbf{x}_{2}) \label{psi1}%
\end{equation}
where $B^{ij}=b^{ik}b^{jk}/2$ is symmetric. Here and throughout the paper, we
imply $\varphi=\varphi\left(  \mathbf{x},t\right)  $ and $\psi=\psi\left(
\mathbf{x},t\right)  ,$ meaning that derivatives are applied to the first two
arguments of the functions. The functional arguments may be omitted when the
context makes this clear. The forward (\ref{fi1}) and backward (\ref{psi1})
equations are formulated for the direct-time Markov diffusion process
$\mathbf{x}_{t}$ (\ref{sde-If}), have adjoint operators on their right-hand
sides, and should not be confused with the corresponding equations for the
reverse-time Markov diffusion process $\mathbf{\bar{x}}_{t}(t)=\mathbf{\bar
{x}}_{\bar{t}}(-t)$. The reverse-time Markov diffusion process $\mathbf{\bar
{x}}_{t}$ defined by Ito SDE (\ref{sde-Ib}) has Kolmogorov forward and
backward equations for corresponding transitional probabilities $\bar{\varphi
}=\bar{\varphi}\left(  \mathbf{x},t\right)  =\bar{\varphi}\left(
\mathbf{x},t;\mathbf{x}_{2},t_{2}\right)  =\mathbb{P}\left(  \mathbf{\bar{x}%
},t|\mathbf{\bar{x}}_{2},t_{2}\right)  $ and $\bar{\psi}=\bar{\psi}\left(
\mathbf{x},t\right)  =\bar{\psi}\left(  \mathbf{x},t;\mathbf{x}_{1}%
,t_{1}\right)  =\mathbb{P}\left(  \mathbf{\bar{x}}_{1},t_{1}|\mathbf{\bar{x}%
},t\right)  $ given by
\begin{equation}
-\frac{\partial\bar{\varphi}}{\partial t}=-\nabla_{i}\left(  \bar{A}^{i}%
\bar{\varphi}\right)  +\nabla_{i}\nabla_{j}\left(  \bar{B}^{ij}\bar{\varphi
}\right)  ,\ \ \text{with}\ \ \left(  \bar{\varphi}\right)  _{t=t_{2}}%
=\delta(\mathbf{x}-\mathbf{x}_{2}) \label{fi2}%
\end{equation}%
\begin{equation}
\frac{\partial\bar{\psi}}{\partial t}=\bar{A}^{i}\nabla_{i}\bar{\psi}+\bar
{B}^{ij}\nabla_{i}\nabla_{j}\bar{\psi},\ \ \text{with}\ \ \left(  \bar{\psi
}\right)  _{t=t_{1}}=\delta(\mathbf{x}-\mathbf{x}_{1}) \label{psi2}%
\end{equation}
where $\bar{B}^{ij}=\bar{b}^{ik}\bar{b}^{jk}/2$. These equations are generally
different from those for the direct-time process $\mathbf{x}_{t}.$

In the Stratanovich interpretation, the Kolmogorov equations are transformed
into%
\begin{equation}
\frac{\partial\varphi}{\partial t}=-\nabla_{i}\left(  a^{i}\varphi\right)
+\nabla_{i}\left(  \frac{b^{ik}}{2}\nabla_{j}\left(  b^{jk}\varphi\right)
\right)  ,\text{ \ \ \ }-\frac{\partial\psi}{\partial t}=a^{i}\nabla_{i}%
\psi+b^{ik}\nabla_{i}\left(  \frac{b^{jk}}{2}\nabla_{j}\psi\right)
\label{fipsi1}%
\end{equation}
for the direct-time process $\mathbf{x}_{t}$ and into%
\begin{equation}
-\frac{\partial\bar{\varphi}}{\partial t}=-\nabla_{i}\left(  \bar{a}^{i}%
\bar{\varphi}\right)  +\nabla_{i}\left(  \frac{\bar{b}^{ik}}{2}\nabla
_{j}\left(  \bar{b}^{jk}\bar{\varphi}\right)  \right)  ,\text{\ \ \ \ }%
\frac{\partial\bar{\psi}}{\partial t}=\bar{a}^{i}\nabla_{i}\bar{\psi}+\bar
{b}^{ik}\nabla_{i}\left(  \frac{\bar{b}^{jk}}{2}\nabla_{j}\bar{\psi}\right)
\label{fipsi2}%
\end{equation}
for the reverse-time process $\mathbf{\bar{x}}_{t},$ where we substitute the
definitions of $A^{i},$ $B^{ij},$ $\bar{A}^{i}$ and$\ \bar{B}^{ij}$. The
boundary conditions imposed on $\varphi,$ $\psi,$ $\bar{\varphi}$ and
$\bar{\psi}$\ remain the same as in \mbox{Equations~(\ref{fi1})--(\ref{psi2})}.

Note that according to the definitions of the transitional probabilities
$\varphi,$ $\psi,$ $\bar{\varphi}$\ and $\bar{\psi}$ require that for any
$t_{1}<t_{2}$
\begin{equation}
\varphi\left(  \mathbf{x}_{2},t_{2};\mathbf{x}_{1},t_{1}\right)  =\psi\left(
\mathbf{x}_{1},t_{1};\mathbf{x}_{2},t_{2}\right)  ,\ \ \ \bar{\varphi}\left(
\mathbf{x}_{1},t_{1};\mathbf{x}_{2},t_{2}\right)  =\bar{\psi}\left(
\mathbf{x}_{2},t_{2};\mathbf{x}_{1},t_{1}\right)  \label{pppp}%
\end{equation}
{While} Kolmogorov forward and backward equations specifically govern
transitional probabilities of a Markov diffusion process, the Fokker--Planck
equation governs diffusive evolution of a more general class of probabilistic
distributions. The forms of the Fokker--Planck and Kolmogorov forward equations
coincide and these terms are often used synonymously, but there still remains
a subtle difference between them. While $\varphi$ and $\psi,$ $\bar{\varphi}$
and $\bar{\psi}$ represent transitional probabilities, physical interpretation
of equations involving forward and backward parabolicities may differ in
different applications. For example, in Conditional Moment Closure
\cite{Klimenko99}, (\ref{fi2}) corresponds to a probability density function,
while (\ref{psi2}) governs the conditional expectation of a scalar.

\subsection{Markov Diffusion Bridge}

At this stage, we turn from consideration of Markov diffusion families,
semigroups, and transitional probabilities to specific processes that are
subject to fixed initial and/or final conditions. While realisations of Wiener
process are time-symmetric---Wiener process viewed in reverse time is still
an equivalent Wiener process-- processes $\mathbf{x}_{t}$ and
$\mathbf{\bar{x}}_{t}$ may or may not have similar time-symmetric statistics
even if the temporal boundary conditions imposed on them are time-symmetric.
The term \textit{bridge} is conventionally used to emphasise setting two
conditions---initial and final---instead of the common preference for
initial conditions only, which is usually justified by antecedent causality.
We impose the same temporal boundary conditions on both $\mathbf{x}_{t}$ and
$\mathbf{\bar{x}}_{t}$
\begin{align}
(\mathbf{\bar{x}}_{t})_{t=t_{1}}  &  =(\mathbf{x}_{t})_{t=t_{1}}%
=\mathbf{x}_{1}\label{bc-i}\\
(\mathbf{\bar{x}}_{t})_{t=t_{2}}  &  =(\mathbf{x}_{t})_{t=t_{2}}%
=\mathbf{x}_{2} \label{bc-f}%
\end{align}
{From} now on, $\mathbf{x}_{1},t_{1},\mathbf{x}_{2}t$ and $t_{2}$\ are fixed
parameters. At this point, we define a new function $f$ that corresponds to the
bridge-conditioned probability $f\left(  \mathbf{x},t;\mathbf{x}_{1}%
,t_{1};\mathbf{x}_{2},t_{2}\right)  =\mathbb{P}\left(  \mathbf{x}%
,t|\mathbf{x}_{1},t_{1};\mathbf{x}_{2},t_{2}\right)  ,$ so that Equation
(\ref{PP-f}) takes the form
\begin{equation}
f\left(  \mathbf{x},t;\mathbf{x}_{1},t_{1};\mathbf{x}_{2},t_{2}\right)
=\frac{\psi\left(  \mathbf{x},t;\mathbf{x}_{2},t_{2}\right)  \varphi\left(
\mathbf{x},t;\mathbf{x}_{1},t_{1}\right)  }{C\left(  \mathbf{x}_{2}%
,t_{2};\mathbf{x}_{1},t_{1}\right)  } \label{fdef-f}%
\end{equation}
where $C$ is defined by $C\left(  \mathbf{x}_{2},t_{2};\mathbf{x}_{1}%
,t_{1}\right)  =\mathbb{P}\left(  \mathbf{x}_{2},t_{2}|\mathbf{x}_{1}%
,t_{1}\right)  ,$ that is $C=\varphi\left(  \mathbf{x}_{2},t_{2}%
;\mathbf{x}_{1},t_{1}\right)  =\psi\left(  \mathbf{x}_{1},t_{1};\mathbf{x}%
_{2},t_{2}\right)  $. For the distribution of the reverse process $\bar
{f}\left(  \mathbf{x},t;\mathbf{x}_{1},t_{1};\mathbf{x}_{2},t_{2}\right)
=$ $\mathbb{P}\left(  \mathbf{\bar{x}},t|\mathbf{\bar{x}}_{1},t_{1}%
;\mathbf{\bar{x}}_{2},t_{2}\right)  ,$ Equation (\ref{PP-b}) takes the form
\begin{equation}
\bar{f}\left(  \mathbf{x},t;\mathbf{x}_{1},t_{1};\mathbf{x}_{2},t_{2}\right)
=\frac{\bar{\psi}\left(  \mathbf{x},t;\mathbf{x}_{1},t_{1}\right)
\bar{\varphi}\left(  \mathbf{x},t;\mathbf{x}_{2},t_{2}\right)  }{\bar
{C}\left(  \mathbf{x}_{1},t_{1};\mathbf{x}_{2},t_{2}\right)  } \label{fdef-b}%
\end{equation}
where $\bar{C}$ is defined by $\bar{C}\left(  \mathbf{x}_{1},t_{1}%
;\mathbf{x}_{2},t_{2}\right)  =\mathbb{P}\left(  \mathbf{\bar{x}}_{1}%
,t_{1}|\mathbf{\bar{x}}_{2},t_{2}\right)  $ (implying $\bar{C}=\bar{\varphi
}\left(  \mathbf{x}_{1},t_{1};\mathbf{x}_{2},t_{2}\right)  =\bar{\psi}\left(
\mathbf{x}_{2},t_{2};\mathbf{x}_{1},t_{1}\right)  $).

The model specified by (\ref{sde-f}) or (\ref{sde-b}) in conjunction with
(\ref{bc-i}) and (\ref{bc-f}) can also be referred to as a \textit{Schr\"{o}dinger
bridge} \cite{Schrodinger1931}, which usually refers to a broader range of
problems associated with finding probability density $\rho(\mathbf{x},t)$ that
is compliant 
with the initial and terminal conditions
\begin{equation}
\rho(\mathbf{x},t)_{t=t_{1}}=\rho_{1}(\mathbf{x)}\text{ and }\rho
(\mathbf{x},t)_{t=t_{2}}=\rho_{2}(\mathbf{x)} \label{SBridge}%
\end{equation}
{The} modern interpretation of Schr\"{o}dinger bridges has evolved towards
broader understanding of the problem including transport optimisation. Given
conditional probability $f\left(  \mathbf{x},t;\mathbf{x}_{1},t_{1}%
;\mathbf{x}_{2},t_{2}\right)  $ and Markov property (\ref{M12}), the solution
of (\ref{SBridge}) can be evaluated by convolution of $f$ with $\rho
_{1}(\mathbf{x)}$ and $\rho_{2}(\mathbf{x)}$, as demonstrated by
Schr\"{o}dinger in their principal work \cite{Schrodinger1931}. While discussion
of a broader spectrum of issues associated with Schr\"{o}dinger bridges is
outside the scope of the present publication, it can be found in the excellent
overview by Chetrite et~al. \cite{S-bridge2021}.

\subsection{Doob's Transform and Anderson's Reversal for the Markov Bridge
\label{appA5}}

In this section, we assume that the processes $\mathbf{x}(t)$ and
$\mathbf{\bar{x}}(t)$ have the same drift and diffusion coefficients as
specified by antisymmetric relations in (\ref{aabb}) in conjunction with the
Stratanovich interpretation $\gamma=\bar{\gamma}=1/2$ of the SDEs. The question becomes whether this ensures that the processes $\mathbf{x}_{t}$ and $\mathbf{\bar{x}}_{t},$ which
are subject to the same boundary conditions (\ref{bc-i}) and (\ref{bc-f}), are
equivalent, and, therefore, have the same distributions $f$ and $\bar{f}$. The
answer is generally negative. Indeed, let us assume that $f=\bar{f}$ is a
smooth solution in the domain $t_{1}\leq t\leq t_{2}$ and create a small
variation $\delta\mathbf{a}$ of drift $\mathbf{a}$ mostly concentrated in a
small vicinity of a selected point $\mathbf{x}_{0},t_{0},$ within the domain
(i.e., $t_{1}<t_{0}<t_{2}$) so that $\left\vert \nabla\cdot\delta
\mathbf{a}\right\vert \gg\left\vert \delta\mathbf{a}\right\vert $. This
alteration would affect functions $\varphi$ and $\bar{\varphi}$ much more than
$\psi$ and $\bar{\psi}$, since the former but not the latter are governed by
conservative equations. These local disturbances $\delta\varphi=\delta
\bar{\varphi}$ are created by the same drift variation preserving the
similarity of the overall distributions $\delta f=\delta\bar{f}$. These
disturbances, however, would diffuse downstream $t>t_{0}$ from $\mathbf{x}%
_{0},t_{0}$ for $\delta\varphi$ and upstream $t<t_{0}$ from $\mathbf{x}%
_{0},t_{0}$ for $\delta\bar{\varphi}$ violating the similarity of $f$ and
$\bar{f}$.\ If $f\neq$ $\bar{f}$, then the direct-time and reverse-time
processes are generally different and require selection of a preferred time
direction to describe the laws of nature. Having a preferred direction of time
corresponds to some form of antecedent causality.

According to (\ref{fi1}) and (\ref{psi1}), $f=f\left(  \mathbf{x},t\right)  $
satisfies the following equation
\begin{equation}
\frac{\partial f}{\partial t}+\nabla_{i}\left(  a^{i}f\right)  =\nabla
_{i}\left(  \frac{b^{ik}\psi}{2C}\nabla_{j}\left(  b^{jk}\varphi\right)
-\frac{b^{ik}b^{jk}\varphi}{2C}\nabla_{j}\psi\right)  =\nabla_{i}\left(
\frac{b^{ik}}{2}\nabla_{j}\left(  b^{jk}f\right)  -\alpha_{^{\text{D}}}^{i}f\right)
\label{f-f}%
\end{equation}
where
\begin{equation}
a_{^{\text{D}}}^{i}=a^{i}+\alpha_{^{\text{D}}}^{i},\ \ \ \alpha_{^{\text{D}}}^{i}=\frac{b^{ik}b^{jk}}%
{\psi}\nabla_{j}\psi=b^{ik}b^{jk}\nabla_{j}\ln\psi=2B^{ij}\nabla_{j}\ln
\psi\label{alf-f}%
\end{equation}
and
\begin{equation}
\text{(a) }\left(  f\right)  _{t=t_{1}}=\delta(\mathbf{x}-\mathbf{x}_{1})\text{
\ \ while\ (b) \ }\left(  f\right)  _{t=t_{2}}=\left(  \psi\right)  _{t=t_{2}%
}=\delta(\mathbf{x}-\mathbf{x}_{2}) \label{tc-f}%
\end{equation}
{Equation} (\ref{f-f}) is interpreted as a direct-time \textit{{Kolmogorov
forward (Fokker--Planck)}} \textit{{equation}} with \textit{{Stratanovich}}
diffusion $b^{jk}$ and drift $a_{^{\text{D}}}^{i}=a^{i}+\alpha_{^{\text{D}}}^{i}$
coefficients, where $f$ is the transitional probability from the state
$\mathbf{x}_{t}(t_{1})=\mathbf{x}_{1}$ to the state $\mathbf{x}_{t}(t).$ While
imposing the initial {condition~(\ref{tc-f}) (a)} 
 on $f$ remains essential, the
final {condition (\ref{tc-f}) (b)}  
 is automatically satisfied due to the
additional drift component $\alpha_{^{\text{D}}}^{i},$ which represents \textit{Doob's
correction drift}. Note that, due to absence of time reversal, this correction
is the same for both Stratanovich and Ito formulations. Equation
(\ref{f-f}) implements \textit{Doob's h-transform} \cite{Doob1957} in
direct Kolmogorov equation, where the conditioning function $h=\psi$ is used
to enforce the final {conditions (\ref{tc-f}) (b)} upon $f$.

Now, we turn to consider the reverse-time process $\mathbf{\bar{x}}%
_{t}$ and its distribution function \ $\bar{f}=\bar{f}\left(  \mathbf{x}%
,t\right)$, which, according to (\ref{fi2}) and (\ref{psi2}), satisfies the
following equation
\begin{equation}
\frac{\partial\bar{f}}{\partial t}+\nabla_{i}\left(  a^{i}\bar{f}\right)
=\nabla_{i}\left(  \frac{b^{ik}b^{jk}\bar{\varphi}}{2\bar{C}}\nabla_{j}%
\bar{\psi}-\frac{\bar{\psi}b^{ik}}{2\bar{C}}\nabla_{j}\left(  b^{jk}%
\bar{\varphi}\right)  \right)  =\nabla_{i}\left(  \frac{b^{ik}}{2}\nabla
_{j}\left(  b^{jk}\bar{f}\right)  -\alpha_{^{\text{A}}}^{i}\bar{f}\right)  \label{f-b}%
\end{equation}
where
\begin{equation}
a_{^{\text{A}}}^{i}=a^{i}+\alpha_{^{\text{A}}}^{i}=-\bar{a}^{i}+\alpha_{^{\text{A}}}%
^{i},\ \ \ \alpha_{^{\text{A}}}^{i}=\frac{b^{ik}}{\bar{\varphi}}\nabla_{j}\left(
b^{jk}\bar{\varphi}\right)  =b^{ik}\nabla_{j}b^{jk}+b^{ik}b^{jk}\nabla_{j}%
\ln\bar{\varphi} \label{alf-b}%
\end{equation}
and
\begin{equation}
\text{(a) }\left(  \bar{f}\right)  _{t=t_{1}}=\delta(\mathbf{x}-\mathbf{x}%
_{1})\text{ \ \ while \ \ (b) }\left(  \bar{f}\right)  _{t=t_{2}}=\left(
\bar{\varphi}\right)  _{t=t_{2}}=\delta(\mathbf{x}-\mathbf{x}_{2})
\label{tc-b}%
\end{equation}
{For} the sake of comparison with (\ref{f-f}), Equation (\ref{f-b})
characterising the reverse-time process is written as direct-time
Fokker--Planck with Stratanovich diffusion $b^{ik}$ and drift $a_{^{\text{A}}%
}^{i}=a^{i}+\alpha_{^{\text{A}}}^{i}$ coefficients and, therefore, involves a time
reversal. The additional drift $\alpha_{^{\text{A}}}^{i}$ specified by (\ref{alf-b})
is \textit{Stratanovich} form of the \textit{Anderson's correction drift}.
This correction ensures a consistent antisymmetric time reversal of a Markov
diffusion process \cite{Anderson1982}, although here we reverse the
reverse-time process $\mathbf{\bar{x}}_{t}$ to run it forward in time---let
us denote this process $\mathbf{\bar{\bar{x}}}_{t}.$ We emphasise that this
reversal is not only a reversal of the Fokker--Planck equation for $\bar{f}$ that
is consistent with the behaviour of $\bar{\varphi}$ at the final state, but
also time reversal of the underlying stochastic process.\ According to the
\textit{Anderson theorem} \cite{Anderson1982}, which goes beyond an earlier
result by Stratanovich \cite{Stratonovich1960}, processes $\mathbf{\bar{x}}%
_{t}$ and $\mathbf{\bar{\bar{x}}}_{t}$ are equivalent---this theorem
guaranties not only probabilistic equivalence of the processes obtained by
Anderson's reversal but, under specified conditions, also does this in a
strong sense implying equivalence of the original and time-reversed stochastic trajectories.

Equation (\ref{f-b}) governs probability distribution $\bar{f}$ for both
processes $\mathbf{\bar{x}}_{t}$ and $\mathbf{\bar{\bar{x}}}_{t}$, so that
$\bar{f}$ automatically satisfies the final {condition (\ref{tc-b}) (b).}  
 Due to
the initial {condition (\ref{tc-b}) (a)}, this $\bar{f}$ can be interpreted as the
transitional probability from the state $\mathbf{\bar{\bar{x}}}_{t}%
(t_{1})=\mathbf{x}_{1}$ to the state $\mathbf{\bar{\bar{x}}}_{t}(t)$ and
(\ref{f-b}) as the corresponding direct-time \textit{Kolmogorov forward
equation}. While processes $\mathbf{\bar{x}}_{t}$ and $\mathbf{\bar{\bar{x}}%
}_{t}$ are equivalent, this is not necessarily the case for $\mathbf{\bar{x}%
}_{t}$ and $\mathbf{x}_{t}$---these processes have not been converted using
Anderson's reversal and,\ indeed, Equations (\ref{f-f}) and (\ref{f-b}) can
have different drifts $a_{^{\text{A}}}^{i}\neq a_{^{\text{D}}}^{i}$ when
$\alpha_{^{\text{D}}}^{i}\neq\alpha_{^{\text{A}}}^{i}$.

The expressions for Anderson's drift $\alpha_{^{\text{A}}}^{i}$ and Doob's drift
$\alpha_{^{\text{D}}}^{i}$ are similar but not the same---these two approaches enforce the
same final conditions but do this differently, with and without the reversal
of time. If, however, $\alpha_{^{\text{D}}}^{i}=\alpha_{^{\text{A}}}^{i},$ then the Kolmogorov
forward Equations (\ref{f-f}) and (\ref{f-b}) have exactly the same diffusion
$b^{ik}$ and drift $a^{i}+\alpha_{^{\text{D}}}^{i}$ coefficients and the same initial
conditions, then the underlying stochastic processes $\mathbf{x}_{t}$ and
$\mathbf{\bar{x}}_{t}$ are also equivalent. Therefore, $f=\bar{f}$ when
$\alpha_{^{\text{D}}}^{i}=\alpha_{^{\text{A}}}^{i}$ so that:
\begin{equation}
\frac{b^{ik}b^{jk}}{\psi}\nabla_{j}\psi=\frac{b^{ik}}{\bar{\varphi}}\nabla
_{j}\left(  b^{jk}\bar{\varphi}\right)  \label{cnd1}%
\end{equation}
{While} Equations (\ref{f-f}) and (\ref{f-b}) are formulated to evolve forward
in time, we could have selected the reverse time $\bar{t}$ for comparison of
$f$ and $\bar{f}$---our formulation is fully time-symmetric. This would lead
to the complimentary condition
\begin{equation}
\frac{b^{ik}b^{jk}}{\bar{\psi}}\nabla_{j}\bar{\psi}=\frac{b^{ik}}{\varphi
}\nabla_{j}\left(  b^{jk}\varphi\right)  \label{cnd2}%
\end{equation}
which can also be derived from $f=\bar{f}$ and (\ref{cnd1}).

Our analysis results in the following theorem:

\begin{theorem}
With consistent choice of the diffusion and drift coefficients, the
direct-time and reverse-time formulations of the Markov diffusion bridge
problem are fully time-symmetric and equivalent if and only if Doob's
transform and Anderson's reversal correction drifts coincide. \label{P3}
\end{theorem}

Indeed, Anderson's theorem establishes the fundamental equivalence of the
reverse-time process $\mathbf{\bar{x}}_{\bar{t}}$ and its time reversal
$\mathbf{\bar{\bar{x}}}_{t}$ running forward in time---both processes have
the same distribution $\bar{f}(\mathbf{x},t).$ When Anderson's and Doob's
drifts coincide $\alpha_{^{\text{D}}}^{i}=\alpha_{^{\text{A}}}^{i}$,\ the processes $\mathbf{\bar
{\bar{x}}}_{t}$ and $\mathbf{x}_{t}$ have the same initial condition, and the
same drift $a_{^{\text{D}}}^{i}=a_{^{\text{A}}}^{i}$ and diffusion $b^{ij}$
coefficients in the direct-time Kolmogorov forward equations and, therefore,
must be equivalent. If $\alpha_{^{\text{D}}}^{i}\neq\alpha_{^{\text{A}}}^{i},$ then the processes
have different drift coefficients and, therefore, are not equivalent
irrespective whether $f$ and $\bar{f}$ \ coincide or not. This proves the theorem.

The consistency of drift and diffusion coefficients in direct-time and
reverse-time models specified by (\ref{aabb}) appears to be insufficient to
ensure that Equations (\ref{f-f}) and (\ref{f-b}) are exactly the same due to
possible differences between Doob's and Anderson's correction drifts,
$\alpha_{^{\text{D}}}^{i}$\ and $\alpha_{^{\text{A}}}^{i}$.  The equivalence of Equations
(\ref{f-f}) and (\ref{f-b}) and the corresponding processes can nevertheless
be achieved but requires additional conditions that are discussed in the next subsection.

\subsection{Antisymmetric Time Reversal and Odd Symmetry}

In line with antisymmetric time reversal (\ref{A12}), (\ref{aabb}), and due to
physical reasons discussed in the main text, we restrict our attention to the
case
\begin{equation}
\bar{a}^{i}=-a^{i},\ \ \bar{b}^{ij}=-b^{ij},\ \ \nabla_{i}b^{ik}%
=0,\ \ \nabla_{i}a^{i}=0\ \ \label{cond3}%
\end{equation}
which ensure that
\begin{equation}
\phi(\mathbf{x},t;\mathbf{x}_{1},t_{1})=\bar{\psi}(\mathbf{x},t;\mathbf{x}%
_{1},t_{1}),\ \ \ \psi(\mathbf{x},t;\mathbf{x}_{2},t_{2})=\bar{\phi
}(\mathbf{x},t;\mathbf{x}_{2},t_{2}) \label{cond3a}%
\end{equation}%
\begin{equation}
\bar{f}\left(  \mathbf{x},t;\mathbf{x}_{1},t_{1};\mathbf{x}_{2},t_{2}\right)
=f\left(  \mathbf{x},t;\mathbf{x}_{1},t_{1};\mathbf{x}_{2},t_{2}\right)
,\ \ \bar{C}=C \label{cond3b}%
\end{equation}
These conditions are important for entropy-consistent formulation of the model
that preserves the phase volume and \textit{odd symmetry} implies in this work
that these conditions are satisfied. With conditions (\ref{cond3}), Equations
(\ref{fipsi1}) and (\ref{fipsi2}) are transformed to take the form of the
Markov bridge model with odd symmetry:
\begin{equation}
\frac{\partial\varphi}{\partial t}=\mathbb{L}\varphi,\ \ \ -\frac{\partial
\psi}{\partial t}=\mathbb{\bar{L}}\psi,\ \ \ -\frac{\partial\bar{\varphi}%
}{\partial t}=\mathbb{\bar{L}}\bar{\varphi},\ \ \ \frac{\partial\bar{\psi}%
}{\partial t}=\mathbb{L}\bar{\psi} \label{L1}%
\end{equation}
where new adjoint operators
\begin{equation}
\mathbb{L}[\cdot]=-\nabla_{i}\left(  a^{i}[\cdot]\right)  +\nabla_{i}\left(
B^{ij}\nabla_{j}[\cdot]\right)  \text{ \ and \ }\mathbb{\bar{L}}%
[\cdot]=+\nabla_{i}\left(  a^{i}[\cdot]\right)  +\nabla_{i}\left(
B^{ij}\nabla_{j}[\cdot]\right)  \ \ \label{L2}%
\end{equation}
are introduced. Under these conditions, the dynamic the direct-time and
reverse-time model are fully equivalent. Due to (\ref{pppp}) and
(\ref{cond3a}), this model is complaint with the odd-symmetric analogue
\begin{equation}
\bar{\varphi}\left(  \mathbf{x}_{1},t_{1};\mathbf{x}_{2},t_{2}\right)
=\varphi\left(  \mathbf{x}_{2},t_{2};\mathbf{x}_{1},t_{1}\right)
,\ \ \ \bar{\psi}\left(  \mathbf{x}_{2},t_{2};\mathbf{x}_{1},t_{1}\right)
=\psi\left(  \mathbf{x}_{1},t_{1};\mathbf{x}_{2},t_{2}\right)  \label{adb}%
\end{equation}
of the even-symmetric detailed balance considered in Appendix \ref{appB}. Note that
Equations~(\ref{L1}) and (\ref{L2}) allow only for constant steady-state solutions
(see Section \ref{Sec31}).

The key point is presented in the form of the proposition:

\begin{proposition}
Doob's transform and Anderson's reversal corrections coincide under
odd-symmetric conditions specified in (\ref{cond3}). Therefore, under these
conditions, the Markov bridge problem is fully time-symmetric and the
direct-time and reverse-time formulations of the problem are equivalent.
\label{P3b}
\end{proposition}

Indeed, substitution of the odd-symmetric conditions of (\ref{cond3}) into
Equations (\ref{fipsi1}) and (\ref{fipsi2}) with boundary conditions taken
from (\ref{fi1})--(\ref{psi2}) results in $\bar{\varphi}=\psi$ and $\bar{\psi
}=\varphi,$ which, with the use of $\nabla_{i}b^{ik}=0,$ ensures that
condition (\ref{cnd1}) is satisfied. Therefore, $\alpha_{^{\text{D}}}^{i}=\alpha_{^{\text{A}}}^{i}$ in
(\ref{f-f}) and (\ref{f-b}) and, according to Theorem \ref{P3}, the problem is
time-symmetric, implying that solving the Markov bridge problem for
$\mathbf{x}_{t}$ or for $\mathbf{\bar{x}}_{\bar{t}}$ is mathematically and
physically equivalent.

Note that condition (\ref{cnd1}) represents $n$ scalar equations $i=1,...,n$
(where $n$ is the number of variables) with $n^{2}$ independent input values
in $b^{ik}$ with only two scalar functions\ $\psi$ and $\bar{\varphi}$\ to
satisfy it---in general, Equation (\ref{cnd1}) is overdetermined. While
constrains~(\ref{cond3}) are sufficient for proper antisymmetric
reversibility, are they necessary to ensure that $\alpha_{^{\text{D}}}^{i}=\alpha_{^{\text{A}}}^{i}%
$? The answer is negative and the class of solutions satisfying $\alpha_{^{\text{D}}}
^{i}=\alpha_{^{\text{A}}}^{i}$ is substantially wider. For example, it is easy to see
that if
\begin{equation}
b^{ik}=\hat{b}^{ik}(\mathbf{x},t)\beta(\mathbf{x}),\ \ \ \ \nabla_{i}\hat
{b}^{ik}=0,\ \ \nabla_{i}\frac{a^{i}}{\beta}=0
\end{equation}
then the conditions (\ref{cnd1}) and (\ref{cnd2}) as well as Equations
(\ref{fipsi1}) and (\ref{fipsi2}) are satisfied by
\begin{equation}
\psi=c_{2}\beta\bar{\varphi},\ \ c_{2}=\frac{1}{\beta(\mathbf{x}_{2}%
)},\text{\ \ }\bar{\psi}=c_{1}\beta\varphi,\ \ \ c_{1}=\frac{1}{\beta
(\mathbf{x}_{1})}=c_{2}\frac{\bar{C}}{C}%
\end{equation}




\renewcommand{\theequation}{B\arabic{equation}}
\setcounter{equation}{0}

\section{Kolmogorov's Concept of Symmetric Reversibility\label{appB}}

Kolmogorov \cite{Kolmogoroff1937} considered a \textit{{Markov diffusion
process}} $\mathbf{x}_{t}$ with a steady-state distribution $\eta\left(
\mathbf{x}\right)  $ satisfying the Fokker--Planck equation in (\ref{fi1})
\begin{equation}
0=\frac{\partial\eta}{\partial t}=-\nabla_{i}\left(  A^{i}\left(
\mathbf{x}\right)  \eta\right)  +\nabla_{i}\nabla_{j}\left(  B^{ij}\left(
\mathbf{x}\right)  \eta\right)  \label{stf}%
\end{equation}
under conditions when the drift and diffusion coefficients $A^{i}\left(
\mathbf{x}\right)  $ and $B^{ij}\left(  \mathbf{x}\right)  $ do not depend on time.

\subsection{Symmetric Time Reversal}

The symmetric time reversal of the process $\mathbf{x}_{t}$ produces a new
process $\mathbf{\bar{x}}_{\bar{t}}$ that that depends on $\bar{t}=-t$ in the
same way that $\mathbf{x}_{t}$ depends on $t$. The reverse-time process can
also be parametrised by $t$ instead of $\bar{t}$ and denoted by $\mathbf{\bar
{x}}_{t}$---this can more convenient for some purposes. The reverse-time
process must also have a steady-state {distribution} 
 $\bar{\eta}\left(
\mathbf{x}\right)  $%
\begin{equation}
0=\frac{\partial\bar{\eta}}{\partial\bar{t}}=-\nabla_{i}\left(  \bar{A}%
^{i}\left(  \mathbf{x}\right)  \bar{\eta}\right)  +\nabla_{i}\nabla_{j}\left(
\bar{B}^{ij}\left(  \mathbf{x}\right)  \bar{\eta}\right)  \label{stb}%
\end{equation}
which, due to the symmetric constrains
\begin{equation}
\bar{A}^{i}\left(  \mathbf{x}\right)  =A^{i}\left(  \mathbf{x}\right)  ,\text{
\ }\bar{B}^{ij}\left(  \mathbf{x}\right)  =B^{ij}\left(  \mathbf{x}\right)
,\ \text{\ }\bar{\eta}\left(  \mathbf{x}\right)  =\eta\left(  \mathbf{x}%
\right)  \label{ABpi}%
\end{equation}
coincides with the steady-state distribution of the direct-time process
$\eta\left(  \mathbf{x}\right)  $.

The similarity between the direct-time and reverse-time processes implies the
following conditions
\begin{equation}
\varphi\left(  \mathbf{x},t;\mathbf{x}_{1},t_{1}\right)  =\bar{\psi}\left(
\mathbf{x}_{1},t;\mathbf{x},t_{1}\right)  ;\ \ \ \ \bar{\varphi}\left(
\mathbf{x},t;\mathbf{x}_{2},t_{2}\right)  =\psi\left(  \mathbf{x}%
_{2},t;\mathbf{x},t_{2}\right)  \label{fi-psi-sym}%
\end{equation}
{Since} the drift and diffusion coefficients are time-independent, the functions
$\varphi,$ $\bar{\psi}$, $\bar{\varphi}$, and $\psi$ depend only on time
differences, i.e., on $t-t_{1}$ or $t_{2}-t$, assuming $t_{1}\leq t\leq t_{2}$
but, in order to avoid confusion, it is more convenient to retain the original
representation of these functions. Indeed, by definition in Section
\ref{sec A-Ke}, $\varphi\left(  \mathbf{x},t;\mathbf{x}_{1},t_{1}\right)
=\mathbb{P}\left(  \mathbf{x},t|\mathbf{x}_{1},t_{1}\right)  $ is the
transition probability from $\mathbf{x}_{t}=\mathbf{x}_{1}$ to $\mathbf{x}%
_{t}=\mathbf{x}$ over time interval $t-t_{1}$, while $\bar{\psi}\left(
\mathbf{x}_{1},t;\mathbf{x},t_{1}\right)  =\mathbb{P}\left(  \mathbf{\bar{x}%
},t_{1}|\mathbf{\bar{x}}_{1},t\right)  $ is the transition probability from
$\mathbf{\bar{x}}_{t}=\mathbf{x}_{1}$ to $\mathbf{\bar{x}}_{t}=\mathbf{x}$
over the same time interval $t-t_{1}$ back in time. These transitional
probabilities must be the same due to equivalence of $\mathbf{x}_{t}$ and
$\mathbf{\bar{x}}_{\bar{t}}$. The second equality in (\ref{fi-psi-sym}) has
the same justification: $\bar{\varphi}\left(  \mathbf{x},t;\mathbf{x}%
_{2},t_{2}\right)  =\mathbb{P}\left(  \mathbf{\bar{x}},t|\mathbf{\bar{x}}%
_{2},t_{2}\right)  $ and $\psi\left(  \mathbf{x}_{2},t;\mathbf{x}%
,t_{2}\right)  =\mathbb{P}\left(  \mathbf{x},t_{2}|\mathbf{x}_{2},t\right)  $.

\subsection{The Detailed Balance Condition}

According to Kolmogorov \cite{Kolmogoroff1937}, the process $\mathbf{x}_{t}$
is called \textit{{reversible}} (i.e., evenly reversible or
\textit{{Kolmogorov-reversible}}) when it satisfies the detailed balance
conditions. Due to equivalence of $\mathbf{x}_{t}$ and $\mathbf{\bar{x}}%
_{\bar{t}},$ the reverse-time process must also satisfy similar conditions.
With the use of new functions
\begin{equation}
\phi(\mathbf{x},t;\mathbf{x}_{1},t_{1})\overset{\text{def}}{=}\frac
{\eta\left(  \mathbf{x}_{1}\right)  }{\eta\left(  \mathbf{x}\right)  }%
\varphi(\mathbf{x},t;\mathbf{x}_{1},t_{1}),\ \ \ \bar{\phi}(\mathbf{x}%
,t;\mathbf{x}_{2},t_{2})\overset{\text{def}}{=}\frac{\eta\left(
\mathbf{x}_{2}\right)  }{\eta\left(  \mathbf{x}\right)  }\bar{\varphi
}(\mathbf{x},t;\mathbf{x}_{2},t_{2}) \label{fi-def}%
\end{equation}
the detailed balance conditions can be expressed as
\begin{equation}
\text{(a) }\phi(\mathbf{x},t;\mathbf{x}_{1},t_{1})=\varphi\left(
\mathbf{x}_{1},t;\mathbf{x},t_{1}\right)  ,\ \ \ \text{(b) }\bar{\phi
}(\mathbf{x},t;\mathbf{x}_{2},t_{2})=\bar{\varphi}\left(  \mathbf{x}%
_{2},t;\mathbf{x},t_{2}\right)  \label{fi-db}%
\end{equation}
{Equation} (\ref{fi-psi-sym}) implies that the detailed balance also requires
that
\begin{equation}
\phi(\mathbf{x},t;\mathbf{x}_{1},t_{1})=\bar{\psi}(\mathbf{x},t;\mathbf{x}%
_{1},t_{1}),\ \ \ \psi(\mathbf{x},t;\mathbf{x}_{2},t_{2})=\bar{\phi
}(\mathbf{x},t;\mathbf{x}_{2},t_{2}) \label{fi-db2}%
\end{equation}
the functions $\phi$ and $\psi$ are, respectively, the same as $\bar{\psi}$ and
$\bar{\phi}$. These conditions are similar to the corresponding conditions
derived for odd symmetry in (\ref{cond3a}). Symmetric time reversal is
considered to have \textit{even symmetry} when it complies with the Kolmogorov
reversibility conditions. Without this compliance, the equalities in
(\ref{fi-db2}) are not satisfied.

The detailed balance conditions are subject to the following reversibility theorem.

\begin{theorem}
(Kolmogorov) Stationary Markov diffusion processes $\mathbf{x}_{t}$ satisfies
the detailed balance {condition (\ref{fi-db}) (a)}  
 (and is, therefore,
Kolmogorov-reversible) when and only when its steady-state distribution
$\eta\left(  \mathbf{x}\right)  $ is compliant with
\begin{equation}
A^{i}\eta=\nabla_{j}\left(  B^{ij}\eta\right)  \label{st2}%
\end{equation}

\end{theorem}

Note that any steady-state solution satisfying (\ref{st2}) also complies with
(\ref{stf}) but not vice versa. The proof of the theorem is given by
Kolmogorov \cite{Kolmogoroff1937}. As the processes $\mathbf{\bar{x}}_{\bar
{t}}$ is a symmetric image of $\mathbf{x}_{t}$, the Kolmogorov reversibility
theorem also applies to $\mathbf{\bar{x}}_{\bar{t}}$ requiring $\bar{A}%
^{i}\bar{\eta}=\nabla_{j}\left(  \bar{B}^{ij}\bar{\eta}\right)  $ for validity
of {(\ref{fi-db}) (b)}.

\begin{proposition}
Anderson's time reversal of stationary Kolmogorov-reversible process
$\mathbf{x}_{t}$ with a steady-state distribution $\eta\left(  \mathbf{x}%
\right)  $ is a stationary reversible process $\mathbf{\bar{x}}_{\bar{t}}$ with
the same steady-state distribution $\bar{\eta}\left(  \mathbf{x}\right)
=\eta\left(  \mathbf{x}\right)  .$
\end{proposition}

Indeed, noting (\ref{st2}) and applying Anderson's time reversal formulae
\cite{Anderson1982} with our present Ito notations, we obtain
\begin{equation}
\bar{B}^{ij}=B^{ij},\ \ \bar{A}^{i}=-A^{i}+2\frac{\nabla_{j}\left(  B^{ij}%
\eta\right)  }{\eta}=-A^{i}+2A^{i}=A^{i}%
\end{equation}
which coincides with (\ref{ABpi}). Under symmetrically steady-state
conditions, Anderson's time reversal, which generally pertains to the
antisymmetric type of reversal of the drift terms $\bar{A}^{i}\sim-A^{i}$, is
flipped by intensive diffusion to comply with the symmetric reversal $\bar
{A}^{i}=A^{i}$. Note that this equation is valid only for steady-state
distributions, while general time-evolving processes $\mathbf{x}_{t}$ and
$\mathbf{\bar{x}}_{\bar{t}}$ with unsteady distributions are not linked by
Anderson's time reversal. \textsl{Also note that the Anderson theorem
\cite{Anderson1982} is by itself a general statement, which is not restricted
to any specific type of symmetry. While implications of the theorem are more
prominent in the case of odd symmetry, Anderson's theorem can also be used in
the case of even symmetry. Yet, the solenoidal odd-symmetric conditions in
(\ref{cond3}) and potential even-symmetric conditions in (\ref{st2}) are
generally incompatible. }

\subsection{Equations for the Transitional Probabilities}

To obtain the even-symmetric form of the equations for transitional
probabilities we substitute $\varphi$ expression in terms of $\phi$ by
(\ref{fi-def}) into (\ref{fi1}) and the coefficient $A^{i}$ determined by
(\ref{st2}) into (\ref{psi1}), take into account that $B^{ij}=B^{ji}$ and
obtain equations for $\phi=\phi(\mathbf{x},t)$ and $\psi=\psi(\mathbf{x},t)$:
\begin{equation}
\eta\left(  \mathbf{x}\right)  \frac{\partial\phi}{\partial t}=\nabla
_{i}\left(  \eta\left(  \mathbf{x}\right)  B^{ij}\left(  \mathbf{x}\right)
\nabla_{j}\phi\right)  ,\ \ \text{with}\ \ \left(  \phi\right)  _{t=t_{1}%
}=\delta(\mathbf{x}-\mathbf{x}_{1}) \label{kfi1}%
\end{equation}%
\begin{equation}
-\eta\left(  \mathbf{x}\right)  \frac{\partial\psi}{\partial t}=\nabla
_{i}\left(  \eta\left(  \mathbf{x}\right)  B^{ij}\left(  \mathbf{x}\right)
\nabla_{j}\psi\right)  ,\ \ \text{with}\ \ \left(  \psi\right)  _{t=t_{2}%
}=\delta(\mathbf{x}-\mathbf{x}_{2}) \label{kpsi1}%
\end{equation}
{With} the use of reversed time $\bar{t}=-t,$ the reverse-time model should look
the same as the direct-time model. The boundary condition for $\phi$ follows
from the boundary condition for $\varphi$\ in (\ref{fi1}) at $\mathbf{x}%
=\mathbf{x}_{1}$. Since we prefer to use $t$\ as time variable, Equations (\ref{fi2}) and~(\ref{psi2}) then become
\begin{equation}
-\eta\left(  \mathbf{x}\right)  \frac{\partial\bar{\phi}}{\partial t}%
=\nabla_{i}\left(  \eta\left(  \mathbf{x}\right)  B^{ij}\left(  \mathbf{x}%
\right)  \nabla_{j}\bar{\phi}\right)  ,\ \ \text{with}\ \ \left(  \bar{\phi
}\right)  _{t=t_{2}}=\delta(\mathbf{x}-\mathbf{x}_{2}) \label{kfi2}%
\end{equation}%
\begin{equation}
\eta\left(  \mathbf{x}\right)  \frac{\partial\bar{\psi}}{\partial t}%
=\nabla_{i}\left(  \eta\left(  \mathbf{x}\right)  B^{ij}\left(  \mathbf{x}%
\right)  \nabla_{j}\bar{\psi}\right)  ,\ \ \text{with}\ \ \left(  \bar{\psi
}\right)  _{t=t_{1}}=\delta(\mathbf{x}-\mathbf{x}_{1}) \label{kpsi2}%
\end{equation}
where $\bar{\phi}=\bar{\phi}(\mathbf{x},t)$ and $\bar{\psi}=\bar{\psi
}(\mathbf{x},t).$ As expected from the Kolmogorov reversibility theorem,
Equations (\ref{kfi1})--(\ref{kpsi2}), which are derived from (\ref{st2}), are
self-adjoint and consistent with~(\ref{fi-db2}), which is equivalent to the
detailed balance~(\ref{fi-db}).

\subsection{Markov Diffusion Bridge Under Even-Symmetric Conditions}

In this section, we consider \textit{Markov diffusion bridge} formulation with
both the initial and final temporal boundary conditions (\ref{bc-i}%
) and (\ref{bc-f}) so that the overall distribution of the direct-time
(\ref{fdef-f}) and reverse-time (\ref{fdef-b}) processes are expressed in
terms of functions $\phi$ and $\psi$ by the formulae
\begin{gather}
f(\mathbf{x},t;\mathbf{x}_{1},t_{1};\mathbf{x}_{2},t_{2})=\frac{\eta
(\mathbf{x})}{C_{1}}\phi\left(  \mathbf{x},t;\mathbf{x}_{1},t_{1}\right)
\psi\left(  \mathbf{x},t;\mathbf{x}_{2},t_{2}\right) \\
\ \bar{f}(\mathbf{x},t;\mathbf{x}_{1},t_{1};\mathbf{x}_{2},t_{2})=\frac
{\eta(\mathbf{x})}{C_{2}}\bar{\phi}\left(  \mathbf{x},t;\mathbf{x}_{2}%
,t_{2}\right)  \bar{\psi}\left(  \mathbf{x},t;\mathbf{x}_{1},t_{1}\right)
\ \ \ \
\end{gather}
with the normalisation constants
\begin{equation}
C_{1}=\eta(\mathbf{x}_{1})\psi\left(  \mathbf{x}_{1},t_{1};\mathbf{x}%
_{2},t_{2}\right)  ,\ \ \ C_{2}=\eta(\mathbf{x}_{2})\bar{\psi}\left(
\mathbf{x}_{2},t_{2};\mathbf{x}_{1},t_{1}\right)
\end{equation}
{Equations} (\ref{kfi1})--(\ref{kpsi2}) indicate that the distributions
$f=f\left(  \mathbf{x},t\right)  $ and $\bar{f}=\bar{f}\left(  \mathbf{x}%
,t\right)  $ satisfy the equations
\begin{equation}
\frac{\partial f}{\partial t}=\nabla_{i}\left(  \frac{\eta B^{ij}}{C_{1}%
}\left(  \psi\nabla_{j}\phi-\phi\nabla_{j}\psi\right)  \right)  =\nabla
_{i}\left(  B^{ij}f\left(  \nabla_{j}\ln\phi-\nabla_{j}\ln\psi\right)
\right)
\end{equation}%
\begin{equation}
\frac{\partial\bar{f}}{\partial t}=\nabla_{i}\left(  \frac{\eta B^{ij}}{C_{2}%
}\left(  \bar{\phi}\nabla_{j}\bar{\psi}-\bar{\psi}\nabla_{j}\bar{\phi}\right)
\right)  =\nabla_{i}\left(  B^{ij}\bar{f}\left(  \nabla_{j}\ln\bar{\psi
}-\nabla_{j}\ln\bar{\phi}\right)  \right)
\end{equation}
which can be transformed into equations
\begin{equation}
\frac{\partial f}{\partial t}=\nabla_{i}\left(  B^{ij}\nabla_{j}%
f\ -\underset{\alpha_{^{\text{D}}}^{i}}{\ \underbrace{B^{ij}\nabla_{j}\ln\left(  \eta
\psi^{2}\right)  }}f\right)  ,\ \ \ \ \ \text{with \ }\left(  f\right)
_{t=t_{1}}=\delta(\mathbf{x}-\mathbf{x}_{1})
\end{equation}%
\begin{equation}
\frac{\partial\bar{f}}{\partial t}=\nabla_{i}\left(  B^{ij}\nabla_{j}\bar
{f}\ -\underset{\alpha_{^{\text{A}}}^{i}}{\ \underbrace{B^{ij}\nabla_{j}\ln\left(
\eta\bar{\phi}^{2}\right)  }}\bar{f}\right)  ,\ \ \ \ \ \text{with \ }\left(
\bar{f}\right)  _{t=t_{1}}=\delta(\mathbf{x}-\mathbf{x}_{1})
\end{equation}
which, as in Appendix \ref{appA5}, are interpreted as direct-time Kolmogorov
forward equations, where the Doob's $\alpha_{^{\text{D}}}^{i}$ and Anderson's $\alpha_{^{\text{A}}}
^{i}$ corrections are the same $\alpha_{^{\text{D}}}^{i}=\alpha_{^{\text{A}}}^{i}$ due to
(\ref{fi-db2}). Hence, Theorem \ref{P3} ensures the following proposition.

\begin{proposition}
Doob's transform and Anderson's reversal corrections coincide under
even-symmetric conditions, which imply Kolmogorov reversibility. Therefore,
the Markov bridge problem is fully time-symmetric under these conditions,
making the direct-time and reverse-time formulations of the problem equivalent.
\end{proposition}

Note that\ if the processes are not Kolmogorov-reversible, the direct-time
$\mathbf{x}_{t}$ and reverse-time $\mathbf{\bar{x}}_{\bar{t}}$ formulations of
the Markov bridge problem are generally not equivalent, even if the equations
in (\ref{ABpi}) are satisfied.

\bibliographystyle{unsrt}
\bibliography{Law3}
\

\end{document}